\documentclass[usenatbib,useAMS]{mn2e}

\usepackage{graphicx}
\usepackage{amsmath}
\usepackage{natbib}
\usepackage{pifont}
\usepackage{txfonts}
\usepackage{mathrsfs}
\usepackage{epsfig}
\usepackage{color}

\newcommand\araa{{ARA\&A}}%
\newcommand\apj{{ApJ}}%
\newcommand\apjl{{ApJ}}%
\newcommand\apjs{{ApJS}}%
\newcommand\aap{{A\&A}}%
\newcommand\icarus{{Icarus}}%
\newcommand\mnras{{MNRAS}}%
\newcommand\ssr{{Space~Sci.~Rev.}}%
\newcommand\jcp{{J.~Chem.~Phys.}}%
\newcommand\planss{{Planet.~Space~Sci.}}%
\newcommand{\sed}[1]{\sed{ \textcolor{blue}{ #1}}}

\renewcommand{\sed}[1]{#1}

\title{Small hydrocarbon molecules in cloud-forming  Brown Dwarf and giant gas planet atmospheres}
\author[C. Bilger, P. Rimmer, Ch. Helling]
       {C. Bilger, P. Rimmer, Ch. Helling\\
        SUPA, School of Physics and Astronomy, University of St. Andrews, North Haugh, St. Andrews, Fife, United Kingdom, KY16 9SS}
\date{Accepted \today}

\pagerange{\pageref{firstpage}--\pageref{lastpage}}
\pubyear{2012}

\begin{document}

\maketitle

\label{firstpage}

\begin{abstract}
\setcounter{page}{1}

We study the abundances of complex carbon-bearing molecules in the
oxygen-rich dust-forming atmospheres of Brown Dwarfs and giant gas
planets. \sed{ The inner atmospheric regions that form the inner 
boundary for thermochemical gas-phase models are investigated.} 
Results from {\sc Drift-phoenix} atmosphere simulations, which
include the feedback of phase-non-equilibrium dust cloud formation on
the atmospheric structure and the gas-phase abundances, are utilised. The resulting
element depletion leads to a shift in the carbon-to-oxygen ratio such
that several hydrocarbon molecules and cyanopolycyanopolyynene molecules can be
present.  An increase in surface gravity and/or a decrease in
metallicity support the increase in the partial pressures of these
species.  CO, CO$_2$, CH$_4$, and HCN contain the largest fraction of
carbon. In the upper atmosphere of low-metallicity objects, more
carbon is contained in C$_4$H than in CO, and also CH$_3$ and
C$_2$H$_2$ play an increasingly important role as carbon-sink.  We
  determine chemical relaxation time-scales to evaluate if hydrocarbon
  molecules can be affected by transport-induced quenching.  Our
  results suggest that a considerable amount of C$_2$H$_6$ and
  C$_2$H$_2$ could be expected in the upper atmospheres not only of
  giant gas planets, but also of Brown Dwarfs. However, the exact
quenching height strongly depends on the data source used. These
results will have an impact on future thermo-kinetic studies, as they change
the inner boundary condition for those simulations.
\noindent
 
\end{abstract} 
\section{Introduction}
\label{sec:intro}

\noindent Brown Dwarfs and giant gas planets have atmospheres in which
carbon is usually less abundant than oxygen.  This may change if the
planet forms in a disk with a carbon-rich dust-gas mixture
\citep[e.g.][]{Fortney2012}, which seems unlikely for Brown Dwarfs
that form from a molecular cloud by gravitational collapse. Both kind
of objects, however, have atmospheres that are so cold that clouds
form from the local atmospheric gas resulting in a depletion of
elements, including oxygen and perhaps carbon.  The formation of mineral
clouds will individually deplete the heavy and less abundant elements
like O, Mg, Si, Fe, Al, Ti \citep{Helling2006,Witte2009} which makes
the observational determination of the element abundances difficult.
The depletion of oxygen is comparably moderate, but strong enough to
shift the carbon-to-oxygen ratio (hereafter C/O ratio) from an initial
solar value of $\sim$0.5 to $\sim$0.7. The hypothesis of carbon-rich
atmospheres in substellar objects is not new, and was inspired by
observations of WASP-12b \citep{Madhusudhan2011} and by questioning
the standard planet composition \citep{Gaidos2000,Seager2007} earlier
on. Our investigations are triggered by our finding
(Sect.~\ref{ss:elmab}) that cloud formation alone causes a
considerable shift in the local C/O ratio and the resulting question
of how this changes the importance of carbon-bearing molecules in
substellar atmospheres.

Independent of the model assumptions in cloud formation is that
methane (CH$_4$) becomes more abundant than carbon monoxide (CO) at
heights above roughly 1 bar in the dense atmospheres of cool Brown
Dwarfs and giant gas planets \citep[][their
  Fig. 4]{Helling2008b}. This is remarkable because the high binding
energy of the CO-molecule causes the blocking of the carbon-chemistry
in an oxygen-rich environment and vice versa. The reason is the high
stability of the CO molecule due to its triple-binding between the C
and the O atoms. CH$_4$ has four single covalent C-H bindings and as a
result, is much more likely to react with other gas-phase species.
Hence, CH$_4$ provides less rigid blocking of the carbon in an
oxygen-rich environment.  We will therefore investigate if this
weakening of the carbon blocking might allow for the presence of even
more complex carbon-binding molecules in the oxygen-rich atmospheres
of Brown Dwarfs and giant gas planets, and this might be affected by
vertical mixing.

 Deviations from local chemical gas-phase equilibrium in the upper
 atmosphere are suggested to be caused by a rapid convective and/or
 diffusive up-mixing of warm gases from deeper atmospheric layers
 \citep[e.g.][]{Saumon2000,Saumon2006,Saumon2007,Leggett2007} in cold
 Brown Dwarfs (e.g. Gliese 229B, 570D) and giant gas planets. A
 similar process is discussed to affect the abundance determination in
 the solar system for Jupiter, Saturn, Uranus, and Neptune
 \citep{Visscher2011}. Another possibility may be the impact of cosmic
 rays on the chemistry of the upper atmospheres of Brown Dwarfs.

Various groups
\citep{Zahnle2009,Zahnle2009b,Line2010,Moses2011,Venot2012} study the
chemical kinetics under the influence of vertical mixing and
photodissociation.  All networks consider molecules formed of the
elements H, C, O and N (\citealt{Line2010} does not include nitrogen).
The number of species and reaction differ between the chemical
networks: \citet{Zahnle2009} (used in \citealt{milric12}) consider 58
species (561 kinetic reactions incl. 33 photo-chemical) and their
network is accurate to C$_2$H$_n$. The species C$_4$H and C$_4$H$_2$
are also included in their network. \citet{Line2010} consider 32
species (299 kinetic reactions incl. 41 photo-chemical), accurate to
C$_2$H$_2$.  \citet{Moses2011} consider 90 species ($\sim$1600
reactions), and incorporate hydrocarbons up to the size of benzine.
\citet{Venot2012} consider 46 C/O species and 42 N-containing species
($>$ 2000 reactions incl. 34 photo-chemical reactions), accurate to
C$_2$H$_n$. All these complex networks prescribe the atmospheric
temperature-pressure profile and treat the eddy diffusion coefficient
K$_{\rm eddy}$ as a parameter. The pressure-dependence of K$_{\rm
  eddy}$ is roughly determined by \citet{Line2010} and
\citet{Moses2011}. \citet{Moses2011} obtains K$_{\rm eddy}$ from the
circulation models of \citet{Showman2009}. \citet{Zahnle2009} treat
the gas temperature as isothermal, and \citet{Moses2011} and
\citet{Venot2012} calculate $(T,p)$ profiles from radiative transfer
\citep{Fortney2006,Fortney2010} and hydrostatic equilibrium
\citep{Showman2009}.  None of these models has a comprehensive
treatment of grain formation or the deposition of gas-phase species
onto grains. The deviation from chemical equilibrium (or a steady
state solution of the kinetic models) increases with increasing mixing
efficiency which is a generally unknown parameter for planetary
  objects outside the solar system \citep{milric12}.  The deviations
also tend to become more significant at lower pressures. The
non-equilibrium steady-state and thermochemical equilibrium abundances
agree to within an order of magnitude when  $p_{\rm gas} \gtrsim 1$bar (=$10^{-6}$ dyn/cm$^2$)  
(see \citealt[their Fig. 8]{Moses2011}.)  The inner boundary is an
additional free parameter the impact of which is demonstrated by
\citet[][their Fig. 1]{Venot2012}. They also point out uncertainties
in the NH$_3$ and HCN abundances based on the use of different
networks.

\sed{ Our ansatz 
considers the collisional dominated inner part of an atmosphere where
kinetic gas-phase modelling is not required, but which composes the
inner boundary for kinetic gas-phase rate networks. In this part of
the atmosphere, the local thermodynamic conditions are well constrained
by atmosphere simulations that combine radiative transfer and cloud
formation (Sect.~\ref{DrPh} for more details). We consider the
part of the atmosphere where the formation of dust clouds influences
the local gas-phase chemistry,  the local temperature and the density. 
The cloud formation causes a depletion of those
elements which take part in the condensation process (e.g. Fe, Mg, O;
Fig.~\ref{fig:dustelem}) resulting in reduced abundances of respective
molecules. Once the cloud particles have formed, they represent a very
strong local  opacity source absorbing in the optical and reemitting
isotropically in the infrared. The consequence is a backwarming effect
which causes a local increase of the gas temperature \sed{ below} the
cloud layer (Fig.~\ref{fig:TP}).

Given that collisional gas-phase processes dominate in the atmospheric
part of interest,} we apply a chemical equilibrium routine that allows
us to provide first estimates of the abundances of carbon-bearing
macro-molecules and small PAHs. This approach allows us to study the
gas-phase abundances at the inner boundary of future kinetic
considerations, and to look at species not presently included in most
of the current networks (e.g. C$_6$H$_6$). Furthermore, we show the
influence of element abundances that are inhomogeneously depleted by
the formation of dust clouds causing, for example, a considerable
shift in the local carbon-to-oxygen (C/O) ratio
(Fig.~\ref{fig:cto}). 
We discuss vertical mixing that transports gas to higher, hence, cooler
atmospheric regions in comparison to the chemical relaxation
timescale, and we assess the influence of uncertainties in rate
coefficient data on the quenching height. The influence of cosmic rays
on the upper atmosphere will be addressed in a forthcoming paper.

\section{Method}
\label{sec:Method}

 We investigate the abundances of complex carbon-binding molecules in
 dust-forming, oxygen-rich atmospheres of Brown Dwarfs and giant gas
 planets by utilising chemical equilibrium calculations \sed{ in
   combination with results from model atmosphere simulations (see Sect.~\ref{DrPh}).}  We
 are particularly interested in studying the effect of a changing C/O
 ratio on the remaining gas-phase chemistry as caused by element
 depletion during cloud formation.


\subsection{Model atmosphere with cloud formation}\label{DrPh}

\begin{figure}
\includegraphics[scale=0.6]{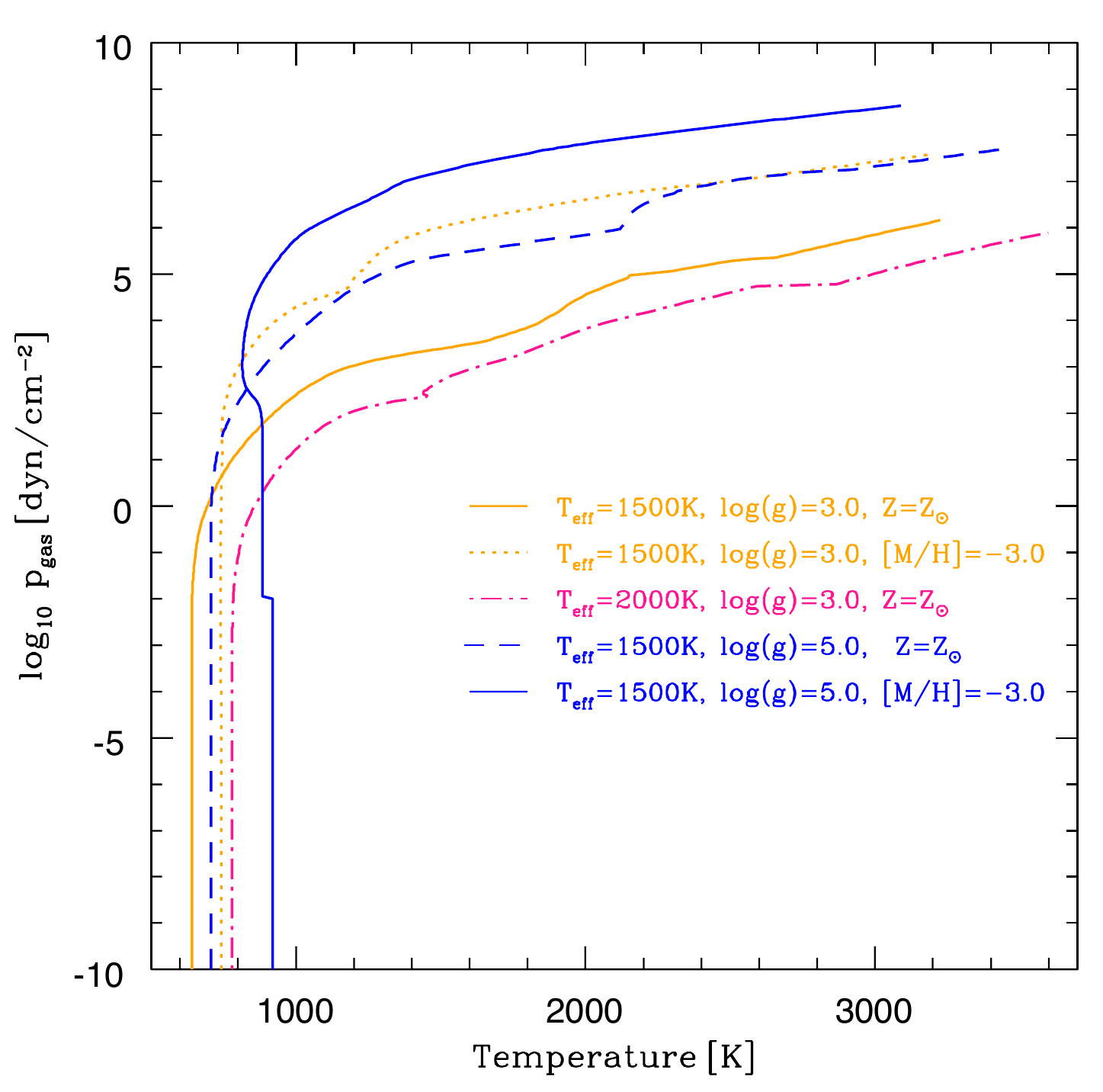}
\caption{ {\sc Drift-Phoenix} model atmospheric structures ($T_{\rm
    gas}$, $p_{\rm gas}$). The low-metallicity models
  ([M/H]$=-3.0$) are always much denser than their solar counterpart
  (Z=Z$_{\odot}$) for a given local temperature.} 
\label{fig:TP}
\end{figure}

\begin{figure}
\includegraphics[scale=0.45]{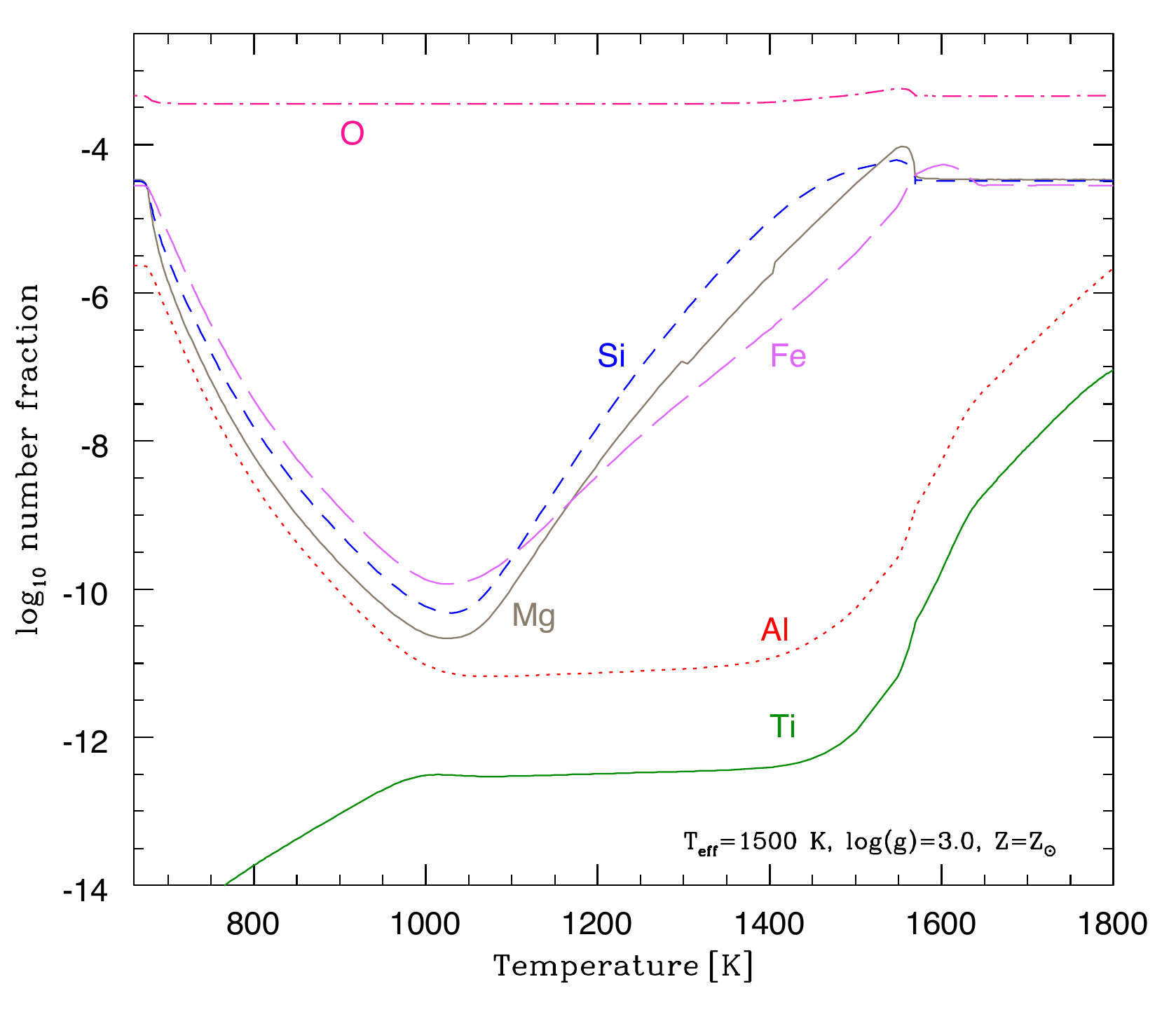}\\*[0.5cm]
\includegraphics[scale=0.45]{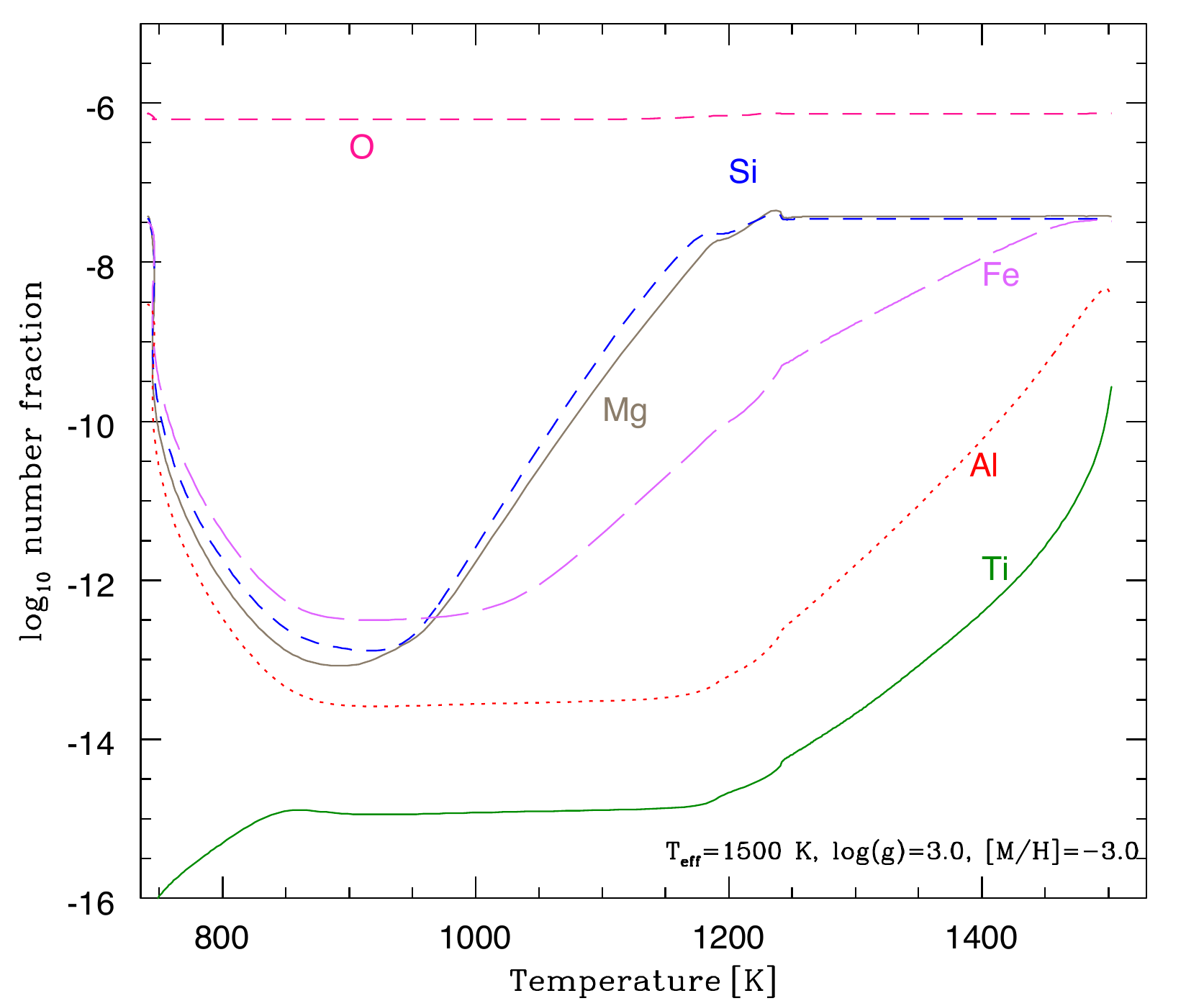}\\*[-0.1cm]
\caption{Gas-phase element abundances affected by dust formation
  in the atmosphere of a giant gas planet ({\sc Drift-phoenix}
  atmosphere model: $\rm T_{\rm eff} = 1500$K, $\log$(g)$ = 3$). \emph{Top}: 
solar composition, \emph{Bottom}: low-metallicity ([M/H]$= -3.0$).}
\label{fig:dustelem}
\end{figure}

\sed{ We utilise results from the {\sc Drift-Phoenix}
  (\citealt{Dehn2007,Helling2008b,Witte2009}) grid of model atmosphere
  simulation which solves the classical 1D model atmosphere problem
  (radiative transfer, mixing length theory, hydrostatic equilibrium,
  gas-phase chemistry; {\sc Phoenix}; Hauschildt \& Baron 1999)
  coupled to a cloud formation model (nucleation, surface growth and
  evaporation, gravitational settling, convective replenishment,  element conservation; {\sc
    Drift}; \citealt{Woitke2003, Woitke2004,Helling2006}). Each of the
  model atmospheres is determined by the effective temperature
  (T$_{\rm eff}$ [K]), the surface gravity (log(g) with g in
  [cm/s$^2$]), and a set of element abundances which have been chosen
  to be solar. These element abundances will be altered where dust
  forms as demonstrated in Fig.~\ref{fig:dustelem}. The metallicity
  may be used as an additional parameter, and can be varied by homogeneously
  increase or decrease all elements to mimic a sub- or supersolar
  element abundance set. Additional input quantities are absorption
  coefficients for all atomic, molecular and dust opacity species
  considered. The cloud's opacity is calculated applying Mie and
  effective medium theory. For more details on {\sf Drift-Phoenix},
  refer to \cite{Witte2009}.

Providing details on the dust clouds, such as height-dependent grain sizes,
and the height-dependent composition of the mixed-material cloud
particles, the model atmosphere code calculates atmospheric
properties, like the local convective velocity, and the spectral energy
distribution, etc. The relevant output quantities which we use for the
present study are the temperature-pressure (T$_{\rm gas}$ [K], p$_{\rm
  gas}$ [dyn/cm$^2$]) structure and the height-dependent element
abundances $\epsilon_{\rm i}$ (Figs.~\ref{fig:TP},~\ref{fig:dustelem}). The local
temperature is the result of the radiative transfer solution and the
local gas pressure of the hydrostatic equilibrium. The element
abundances are the result of the element conservation equations that
include the chance of elements by dust formation and evaporation.

}

\subsection{Chemical equilibrium calculation}
\label{sec:thermochemical-data}

In \sed{ Local Thermodynamic Equilibrium (LTE)}, at a given temperature, gas pressure,
and
for a elemental composition, the chemical abundances depend on the
thermodynamic properties of the species through their pressure
equilibrium constants, K$_p$(T). For  an example dielemental molecule being
 formed from  gaseous constituent atoms, K$_p$(T) is a function
of temperature only, and is given by the law of mass action
\citep[e.g.][]{tsuji73,Gail1986},
\begin{equation}
\ln K_p=\ln\frac{p(\rm A)^{a}p(\rm B)^{b}}{p(\rm A_aB_b)}
\end{equation}
where $p(\rm A_aB_b)$, $p(\rm A)$ and $p(\rm B)$ are the partial pressures of the molecule 
A$_a$B$_b$ in LTE and constituent atoms A and B, respectively.
The temperature dependence of K$_p$(T) can be fitted with a $4^{\rm{th}}$ order polynomial
\begin{equation}
\ln \mathrm{K_p} = a_0 + a_1 \theta + a_2 \theta^2  +a_3 \theta^3  + a_4\theta^4
\end{equation}
where $\theta$ is the reciprocal temperature equal to $5040$/T
(e.g. Tsuji 1973).  Tabulated fitting parameters for
the complex carbon-bearing molecules are from \citet{Cherchneff1992}.
Each atom, molecule and ion is represented by the law of mass action,
as well as satisfying element conservation and charge
conservation. The partial pressure of the molecules, ions and atoms
can then be converted into a number density, $n_{A_aB_b}$,  by use of the ideal gas
law
\begin{equation}
p(A_aB_b) = n_{A_aB_b}kT.
\end{equation}
The equations are solved simultaneously for all gas-phase species.

The atmospheric profile of the local gas temperature, gas pressure,
and element abundances are prescribed ({\sc Drift-Phoenix} model
atmosphere results; \citet{Dehn2007,Witte2009}) and the chemical
equilibrium is evaluated for each atmospheric layer.

\subsection{Chemical kinetic approach through quenching kinetics}
\label{sec:non-equilibrium}



In the deep atmospheric layers, at high temperatures and pressures,
chemical equilibrium can prevail if reaction kinetics operate faster
than convective mixing. \sed{ This is when the time-scale for a species
  to reach thermochemical equilibrium, $t_{\rm chem}$, is less than
  the timescale for atmospheric mixing, $t_{\rm mix}$
  (Sect. \ref{subsec:quenching} for more details)}.  We consider
vertical mixing as the \emph{only} cause of non-equilibrium gas-phase
chemistry here.  We consider  non-irradiated objects only, and hence, 
can neglect the effect of photo-chemistry for the time
being.

\subsubsection{Chemical relaxation time scale}

 Consider, for example, a gas-phase species, A, which is formed and
destroyed solely by the gas-phase reaction:
\begin{equation*}
 A + B \rightleftharpoons C + D.
  \label{eqn:reversible}
 \end{equation*}
 The change in the number density of A, ${\rm [A]}$ [cm$^{-3}$], is:
 \begin{equation*}
  \Bigg(\dfrac{d{\rm[A]}}{dt}\Bigg)_{p,T} = k_r(p,T){\rm[C][D]} - k_f(p,T){\rm[A][B]},
  \label{eqn:basic-kinetics}
 \end{equation*}
 where $k_f(p,T)$ is the forward reaction rate coefficient, and
 $k_r(p,T)$ is the reverse reaction rate coefficient. Consider a given
 pressure and temperature, $(p_0,T_0)$, where
 $(d{\rm[A]_0}/dt)_{p_0,T_0} = 0$.
 The densities ${\rm[A]_0}$,${\rm[B]_0}$, etc. are the equilibrium densities for $(p_0,T_0)$. We now quickly\footnote{
 fast enough that ${\rm [A]} \approx {\rm[A]_0}$, ${\rm [B]} \approx {\rm[B]_0}$, etc.} transport A in a large gas parcel to a new pressure and 
 temperature, $(p_1,T_1)$, with new equilibrium densities,
 ${\rm[A]_1}$,${\rm[B]_1}$, etc., where $(d{\rm[A]_1}/dt)_{p_1,T_1} = 0$. If
${\rm[A]_0} \gg {\rm[A]_1}$, then the time-scale, $t_{\rm chem}$, for A to go from 
 ${\rm [A]_0} \rightarrow {\rm [A]_1}$ can be expressed as \citep{Prinn1977}:
 \begin{align*}
  \dfrac{1}{t_{\rm chem}} &= \dfrac{1}{\rm[A]_0}\Bigg|\dfrac{d{\rm[A]_0}}{dt}\Bigg|_{p_1,T_1} \\
  t_{\rm chem} & = \dfrac{\rm [A]_0}{|k_f(p_1,T_1){\rm[A]_0[B]_0} -k_r(p_1,T_1){\rm[C]_0[D]_0}|}.
 \label{eqn:time-scale-full}
 \end{align*}
The condition, ${\rm [A]_0} \gg {\rm [A]_1}$, requires that:
 \begin{equation*}
 k_f(p_1,T_1){\rm[A]_0[B]_0} \gg k_r(p_1,T_1){\rm[C]_0[D]_0}.
 \label{eqn:fr-inequality}
 \end{equation*}
 If this were not the case, then the formation rate for A would be nearly equal or greater than the destruction rate for A, and
at steady state, we would find ${\rm [A]_0} \lesssim {\rm [A]_1}$, violating our required condition. 
In the case where ${\rm [A]_0} \gg {\rm [A]_1}$, the reverse reaction rate can
 be neglected, and:
 \begin{equation}
  t_{\rm chem} \approx \dfrac{1}{k_f(p_1,T_1){\rm[B]_0} }.
  \label{eqn:time-scale-approx}
 \end{equation}
With this we consider that species A is destroyed by
$N$  reactions, each involving a species B$_i$, governed by a
rate constants $k_i$. In this
case, Eq.~\ref{eqn:time-scale-approx} is generalised  as
\begin{equation}
t_{\rm chem}({\rm A}) = \bigg(\sum_{i=0}^{N} k_i [{\rm B}_i]_0\bigg)^{-1}.
\label{eqn:routes}
\end{equation}
If a particular $k_j [{\rm B}_j]_0$ is much larger than any of the other $k_i [{\rm B}_i]_0$, then 
Eq.~\ref{eqn:routes} simplifies to
\begin{equation}
\label{eq:tchemBj}
t_{\rm chem}({\rm A}) \approx \big(k_j [{\rm B}_j]_0\big)^{-1}.
\end{equation}
The dominating destruction rate depends both on the rate coefficient,
$k_j$, and on the number density of species B, [B$_{j}$]$_{0}$ in the
gas parcel.  The chemical kinetic conversion timescale, $t_{\rm
  chem}({\rm A})$ [s], for a given gas-phase species A, is the time
for relaxation towards an equilibrium state. It is defined here as the
time it takes for the number density, $[{\rm A}]$ [cm$^{-3}$], to
reach the equilibrium value.  We describe our method for finding the
dominant destruction reactions for select hydrocarbons in Section
\ref{ss:nonequapproach}.

\subsubsection{Mixing time scale}
Large-scale convection is the transport of gases with the mean bulk
flow. In contrast, diffusion refers to the transport of gases along a
negative concentration gradient by the action of random motions. In
the radiative zone, the vertical transport timescale by eddy
diffusivity was suggested to be \citep[e.g.][]{Saumon2006,Moses2011}.
\begin{equation}
\label{eq:taueddy}
t_{\rm mix,eddy}=\frac{H_{\rm p}(z)^2}{K_{\rm eddy}},
\end{equation}
where H$_{\rm p}(z)$ is the local pressure scale height.  The
coefficient for eddy diffusion $K_{\rm eddy}$ was taken to be $10^4$
and 10$^8$ $\rm cm^2 s^{-1}$, a set of reasonable values for
substellar atmospheres \citep{Saumon2006, Saumon2007}.  A comparison
with \citet{Zahnle2009} and \citet{Moses2011} show that $K_{\rm eddy}
= 10^8$ $\rm cm^2 s^{-1}$ is on the low end of the values used in
kinetic models. However, \citet{milric12} studies mixing efficiencies
down to $K_{\rm eddy} = 10^6$ $\rm cm^2 s^{-1}$ for GJ1214b.
Increasing the value of $K_{\rm eddy}$ increases the mixing which can
yield larger departures from chemical equilibrium as shown by e.g
\citet{Zahnle2009}.  The convective zone is situated in deeper
atmospheric layers (the convective velocity is $\neq 0$), however,
\citet{Woitke2004} developed the idea of convective overshooting.
Driven by their momentum, the parcels of gas are able to ascend the
atmosphere beyond the Schwarzschild boundary into the radiative
zone. We compare the chemical kinetic conversion timescales ($t_{\rm
  chem}$; coloured lines in Figs.~\ref{fig:C2H2},~\ref{fig:C2H6}) for
the hydrocarbon species to the vertical convective overshooting
timescales (thin black lines in Figs.~\ref{fig:C2H2},~\ref{fig:C2H6})
of each atmosphere \citep[see][their Eq. 9]{Woitke2004} and to the
eddy diffusion timescales (Eq.~\ref{eq:taueddy}, thick black lines in
Figs.~\ref{fig:C2H2},~\ref{fig:C2H6}).  The convective mixing involved
in the Drift-Phoenix models does mainly impact chemistry-wise the
cloud forming part of the atmosphere and decreases exponentially in
the upper atmosphere. The diffusive mixing as applied in most
of the complex chemical network evaluations sustains its efficiency
throughout the whole atmosphere, moving slower in the inner atmosphere{and faster in the higher atmosphere.

\subsubsection{Quenching level}
\label{subsec:quenching}

\sed{ Deep in the atmosphere, the species in a gas parcel reach equilibrium with
the surrounding gas faster than the time it takes for the gas parcel
to reach the upper atmosphere. In the outer cooler atmosphere,
however, energy barriers can become significant and vertical transport
can dominate over chemical processes ($t_{\rm chem} > t_{\rm
  mix}$). As a result, departures from chemical equilibrium can
potentially be observed for some molecules 
  \citep[e.g. CH$_4$/CO,][]{Saumon2006}. The abundance of a molecular constituent may
become ``quenched'' at a value called the ``quench level'', when
$t_{\rm chem} = t_{\rm mix}$
\citep{Prinn1977,Saumon2006,Visscher2011,Moses2011}.  Above that
level, at lower gas temperatures, the chemical reactions are frozen by
vertical mixing, i.e. the forward chemical timescale is significantly
slower than the gas-dynamic timescales involved. The intersection
between the mixing timescale and the reaction timescale of a chemical
species marks the point inside the atmosphere where the equilibrium
number density of that species is ``frozen in'', and from where on it
remains roughly constant.}

\begin{figure}
\hspace*{-0.5cm}
\includegraphics[scale=0.6]{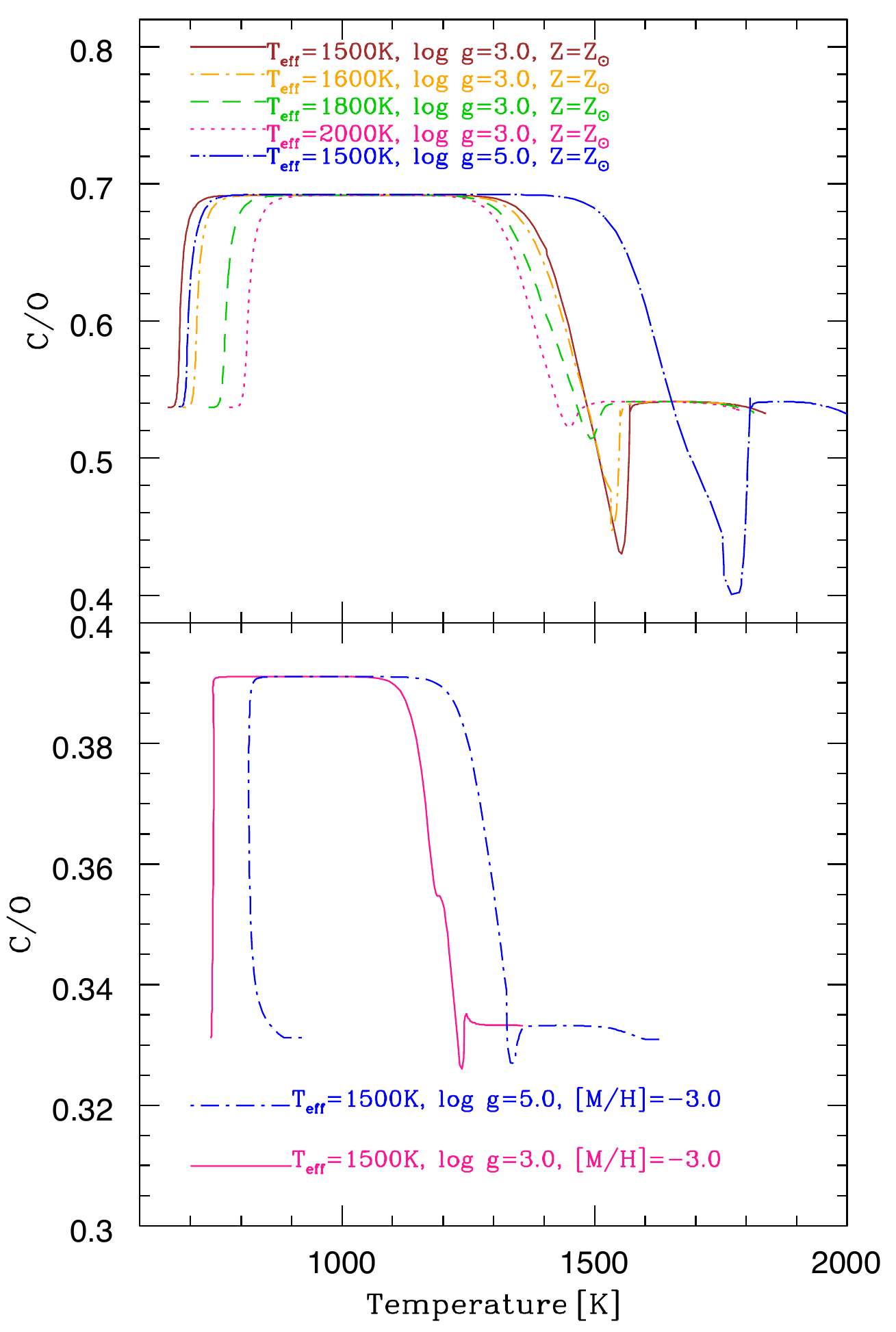}
\caption{Carbon-to-oxygen ratios inside the dust cloud layer depending
  on T$_{\rm eff}$, log(g) and metallicity [M/H]. The decrease below
  the initial value (top: solar, bottom: [M/H]$= -3.0$) of the C/O
  ratio, is caused by the evaporation of the cloud particles.}
\label{fig:cto}
\end{figure}

\section{Approach}
\label{sec:approach}

\subsection{Equilibrium gas-phase chemistry calculation}
\label{sec:equilibrium-chem}

A combination of $199$ gas-phase molecules (including $33$ complex
carbon-bearing molecules), $16$ atoms, and various ionic species were
used under the assumption of LTE. This is an extension of the
gas-phase chemistry routine used so far in our dust cloud formation
according to \citet{Helling1996}.  The data for the large
carbon-bearing molecules considered are taken from
\citet{Cherchneff1992} and they are grouped according to their
structure as follows: large N-bearing species HC$_x$N, complex
hydrocarbons C$_n$H$_{2n\pm2}$, C$_2$H$_{2n}$, CH-bearing radical
C$_x$H, CH$_x$, and C$_x$.  The \citet{Grevesse2007} solar composition
is used for calculating the gas-phase chemistry outside the metal
depleted cloud layers and before cloud formation.
No solid particles were included in the chemical equilibrium
calculations but their presence influences the gas phase by the
reduced element abundances due to cloud formation and the cloud
opacity impact on the radiation field, both accounted for in the {\sc
  Drift-phoenix} model simulations (Sect~\ref{DrPh}).  We utilize {\sc
  Drift-Phoenix} model atmosphere (T$_{\rm gas}$, p$_{\rm gas}$,
$\epsilon_{\rm i}$,) structures as input for our chemistry
calculations.


\subsection{Reaction kinetics and rate determining steps}
\label{ss:nonequapproach}

\begin{table*}
\begin{minipage}[ht]{15cm} 
\caption{Hydrocarbons' chemical reactions for kinetic calculations. 
The molecule M that appears in a few chemical reactions represents any third body. 
$A$, $n$ and $E_a$ are 
used to calculate the rate coefficients $k$ \emph{for forward reactions}, 
although all the reactions are reversible.}
\centering
\begin{tabular}{l l c c c l}
\hline
\hline
Reaction & $A$ [cm$^3$ s$^{-1}$] & $n$ & $E_a$ [K] & $\Delta$T [K] & References\\
\hline
C$_2$H$_2$ + H $\xrightarrow{\rm M}$ C$_2$H$_3$ & $3.3 \times 10^{-30}$ $^a$ & 0.0 & 740 & $200-400$ & \citet{Baulch1992}\\
$^b$ & $1.4 \times 10^{-11}$ & 0.0 & 1220 & $300-2000$ & \citet{Baulch1992}\\
C$_2$H$_2$ $\xrightarrow{\rm M}$ Products & $1.1 \times 10^{-10}$ & 0.0 & 20880 & $1500-3500$ & \citet{Duran1989}\\
$^b$ & $1.0 \times 10^8$ $^c$ & 0.0 & 20400 & $700-2500$ & \citet{Klotz1980}\\
\hline
C$_2$H$_3$ + H$_2$ $\rightarrow$ C$_2$H$_4$ + H & $3.4 \times 10^{-14}$ & 2.6 & 2525 & $300-2500$ & \citet{Knyazev1996}\\
\hline
C$_2$H$_6$ $\xrightarrow{\rm M}$ CH$_3$ + CH$_3$ & $1.9 \times 10^9$ & -13 & 51000 & $700-1924$ & \citet{Oehlschlaeger2005}\\
$^b$ & $1.7 \times 10^{26}$ $^c$ & -9.7 & 53900 & $700-1924$ & \citet{Oehlschlaeger2005}\\
\hline
\end{tabular}
\footnotetext[1]{Units of [cm$^6$ s$^{-1}$]}
\footnotetext[2]{This rate coefficient is for the reaction above in the high pressure limit. This can be combined
with the above rate coefficient as described in \citet{Visscher2011}.}
\footnotetext[3]{Units of [s$^{-1}$]}
 \vspace{-7pt}\renewcommand{\footnoterule}{}
\label{table:reaction}
\end{minipage}
\end{table*}

Various gas-phase rate networks are applied in the literature to study
non-equilibrium gas-phase abundances in the upper low-pressure
planetary atmospheres for irradiated objects
\citep{Zahnle2009,Zahnle2009b,Line2010,Moses2011,Venot2012,milric12,Kopparapu2012}. 
\sed{ Complete chemical networks provide the ideal basis for determining 
quenching heights. The standard approach is to first model the atmosphere of a given object using
a full rate network, and then to analyse the results of this network. Sensitivity analysis of networks
can find rate-determining steps \citep{Moses2011}, and can even guide construction of a simplified
network that includes only the dominant reactions (Carrasco et al. 2008)}. A
comparison of the results \sed{ for these networks} is generally difficult because the rate
networks differ in the number of rates. Therefore, the completeness of
reaction paths, and often the reaction coefficient
\citep[e.g.][]{Venot2012} and mixing parameter \citep[e.g.][]{milric12}
are not well constrained. A further challenge for rate network
calculations is the choice of the inner boundary and the initial
values for their kinetic equations. 

 \sed{ Thus far, we have followed a chemical equilibrium approach to
   gain a first insight of how much element depletion by dust
   formation would support the existence of large carbon molecules in
   the collisional dominated part of the atmosphere which has
   well-determined thermodynamic properties (see Sect~\ref{DrPh})
   compared to the diluted photo-chemically drive exosphere.  To
   assess potential non-equilibrium effects on macroscopic
   carbon-binding molecules in the gas-phase due to quenching,}
  we study time-scale for only C$_2$H$_2$, C$_2$H$_3$ and
 C$_2$H$_6$, for which we can provide the equilibrium values.  We also
 investigate the possible effects of varying model atmosphere
 parameters (e.g. the surface gravity log(g)) \sed{ in order to
   uncover potential differences between the extended atmosphere of a
   planet, and the much more compact atmosphere of Brown Dwarfs.}

To investigate which molecule is driven out of equilibrium,
one ideally uses an entire rate network because minor species might
become unexpectedly important. However, it may also be possible to
identify the dominant rate determining step which then would allow us
an efficient assessment by using Eq.~\ref{eq:tchemBj}. This approach
has been followed e.g. by \citet{Saumon2006} and was addressed also in
\citet{Moses2011}, and it has the advantage of flexibility. It further
allows us to evaluate uncertainties in material quantities efficiently
(Sect.~\ref{ss:uncertain}).  \sed{ Carrasco et al. (2008) demonstrate
  how to produce a skeletal chemistry for Titan with the help of
  reference mass spectrum from INMS measurements of the Cassini
  Spacecraft. Their research is driven by the understanding that
  not all 700 reactions are needed to describe the Cassini
  observations. Their criterion is the reproduction of a certain
  reference observable within a certain uncertainty by a reduced set
  of reactions, starting from the full network.} 

\begin{figure}
\hspace*{-0.5cm}
\includegraphics[scale=0.6]{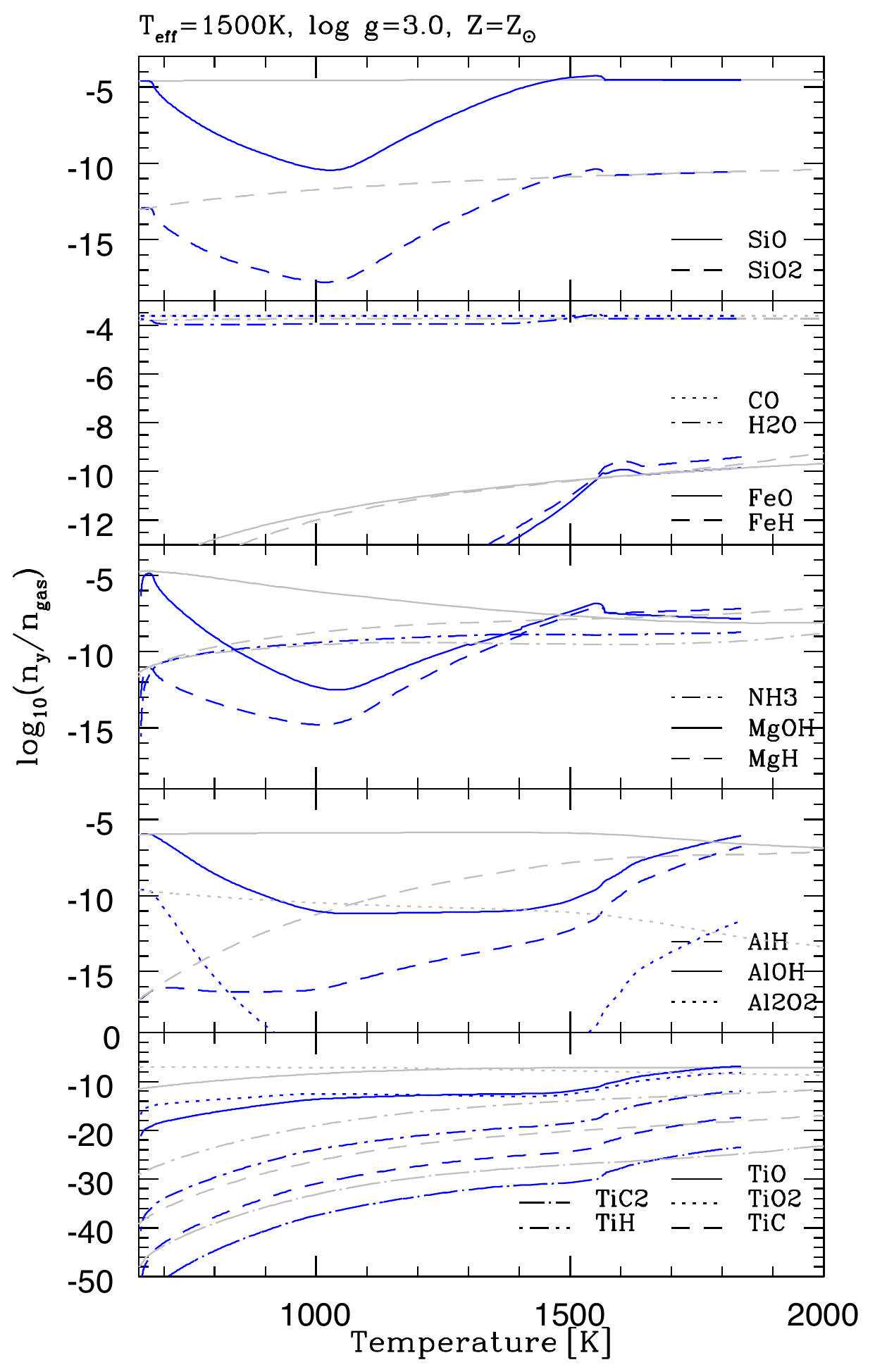}
\caption{The impact of dust-depleted element abundances on the
  gas-phase composition. Only molecules involved in dust formation and
  those typically appearing in a dense, oxygen-rich atmosphere are
  shown: grey -- no element depletion (solar abundances), blue --
  dust depleted element abundances. We utilise a giant gas planet {\sc
    Drift-Phoenix} model atmosphere (T$_{\rm eff}$=1500K, log(g)=3.0,
  initial solar).}
\label{fig:others15003}
\end{figure}

We will explore the quenching heights of C$_2$H$_2$,
C$_2$H$_3$, and C$_2$H$_6$ using a simplified method:
We consider only the immediate destruction reactions for C$_2$H$_2$, C$_2$H$_3$, 
and C$_2$H$_6$ in order to calculate the value of $t_{\rm chem}$ 
using Eq.~\ref{eq:tchemBj}. We search for the dominant destruction reaction over
the entire (T$_{\rm gas}$, p$_{\rm gas}$) profile. Three criteria guide our search for the dominant 
immediate destruction reaction. For a given
reaction involving species A with abundance ${\rm [A]_0}$,
with a destruction reaction frequency $\nu_f$ [s$^{-1}$]\footnote{ \sed{ The reaction frequency is defined
 as $\nu \equiv k_2 [M]$
for combustion reactions, 
$\nu \equiv k_2 {\rm[B]}$ for two-body reactions involving the species B, and 
$\nu \equiv k_3 {\rm[B]}[M]$ for three-body reactions involving B. In each case,
$k_2$ [cm$^3$ s$^{-1}$] and $k_3$ [cm$^6$ s$^{-1}$] are measured rate coefficients
and $[M]$ [cm$^{-3}$] is the number density of the third body.}}, 
and reverse reaction with a frequency $\nu_r$ [s$^{-1}$]$^2$, there may be a range of (T$_{\rm gas}$, p$_{\rm gas}$) where:
\begin{enumerate}
 \item The reaction is reversible.
 \item $\nu_f \gg \nu_r$.
 \item $\nu_f$ is much greater than the measured destruction frequency for other destruction reactions.
\end{enumerate}
If these three criteria are met for a range of our (T$_{\rm gas}$, p$_{\rm gas}$)  profile, then we consider that reaction to be the dominant
destruction reaction for that range.
 We applied these three criteria in the case of C$_2$H$_2$, C$_2$H$_3$, 
and C$_2$H$_6$. We obtained the rate coefficients for the destruction reactions 
by searching the NIST Kinetics Database (http://kinetics.nist.gov/),
the UMIST Database for Astrochemistry, the reaction list from
\citet[][www.lpi.usra.edu/science/moses/reaction\_list.pdf]{Moses2005}
and from \cite[][]{Venot2012}, available on the KIDA database
\citep[http://kida.obs.u-bordeaux1.fr]{Wakelam2012}. Reverse reactions
were obtained from NIST, where available. When NIST did not list the
reverse reactions, the rate coefficients were calculated from the
assumption of microscopic reversibility, employing the method outlined
in \citet{Visscher2011}.  For each of the dozens of reactions, we calculated
the destruction reaction frequency, and the reverse (formation) reaction frequency. The
reactions that met all three criteria above are listed along with their
rate coefficients in Table~\ref{table:reaction}.

 The rate coefficients for the dominant destruction reactions, $k$, are given by 
the Arrhenius equation:
\begin{equation}
k= A \, \Bigg(\frac{T}{\rm 298 \; K}\Bigg)^{n} \; e^{-E_a/T}.
\label{eq:arrhenius}
\end{equation}
Where $A$, $n$ and $E_a$ are parameters taken from the sources mentioned above. 
Units of $k$ are $\rm s^{-1}$, $\rm
cm^3 \; \rm s^{-1}$ and $ \rm cm^6 \; \rm s^{-1}$ for first-, second-
and third-order reactions, respectively. Table \ref{table:reaction} 
contains the  dominant destruction  reactions for
 C$_2$H$_2$, C$_2$H$_3$ and C$_2$H$_6$.
Multiple destruction  rates are listed because
 different destruction reactions dominate at different pressures.

The listed reactions may differ for different 
\sed{ atmospheres of different objects or
for an atmosphere with different element abundances, e.g. for a} 
N$_2$ dominated atmosphere \sed{ which is the case also for full rate networks.} This method cannot 
account for the nonlinear nature of the full kinetics treatment, but no current
chemical kinetics model is able to account for the deposition of
gas-phase species onto dust grains as done in the {\sc Drift-Phoenix} atmosphere 
models used here.

\sed{ The purpose of our approach is 
to constrain quenching heights for select small hydrocarbons to
indicate potentially arising non-equilibrium effects starting from our
equilibrium abundances of carbon-binding macro-molecules.  Since our
intent is not to incorporate any non-equilibrium processes other than
atmospheric mixing, this simplistic time-scale comparison seems
capable of 
 determining whether a given species is
quenched, and what range of heights at which it is likely to be
quenched. Although solving the series of nonlinear coupled
differential equations describing the full chemical kinetics is in
general the 
most complete approach, in this case it would risk amplifying
uncertainties \citep{Wakelam2005} by using large computing resources.
A detailed discussion of the reasons for these uncertainties and their
effect on our investigation is presented in
Sec. \ref{sec:mixing-results}}.

A considerable uncertainty is hidden in the designation of the
reactions in Table \ref{table:reaction} due to uncertainties in
laboratory and theoretical investigations. Errors in the treatment of
reaction kinetics (incorrect assumptions about the rate-determining
step or rate coefficient) are common, and may lead to uncertainties in
the conclusions regarding disequilibrium abundances. This has been
demonstrated by \citet{Venot2012} for the competing abundances of
NH$_3$ and HCN. This problem is known in the wider astronomical
community and has lead to comparative studies of PDR chemical networks
(e.g. R{\"o}llig et al 2007; \sed{ PDR - photo-dissociation regions}).

\sed{ Our list of reactions in Table 1
is constructed based on the best available measured and calculated rate
coefficients. As this knowledge improves, the list of reactions might
change, and so would the quenching height, independent of the approach chosen. 
Similar risks are present when analyzing the chemistry using a comprehensive
chemical network. To our knowledge, the reactions in Table 1 are the 
most efficient destruction routes for the species we consider.}

\begin{figure}
\begin{center}
\includegraphics[scale=0.39]{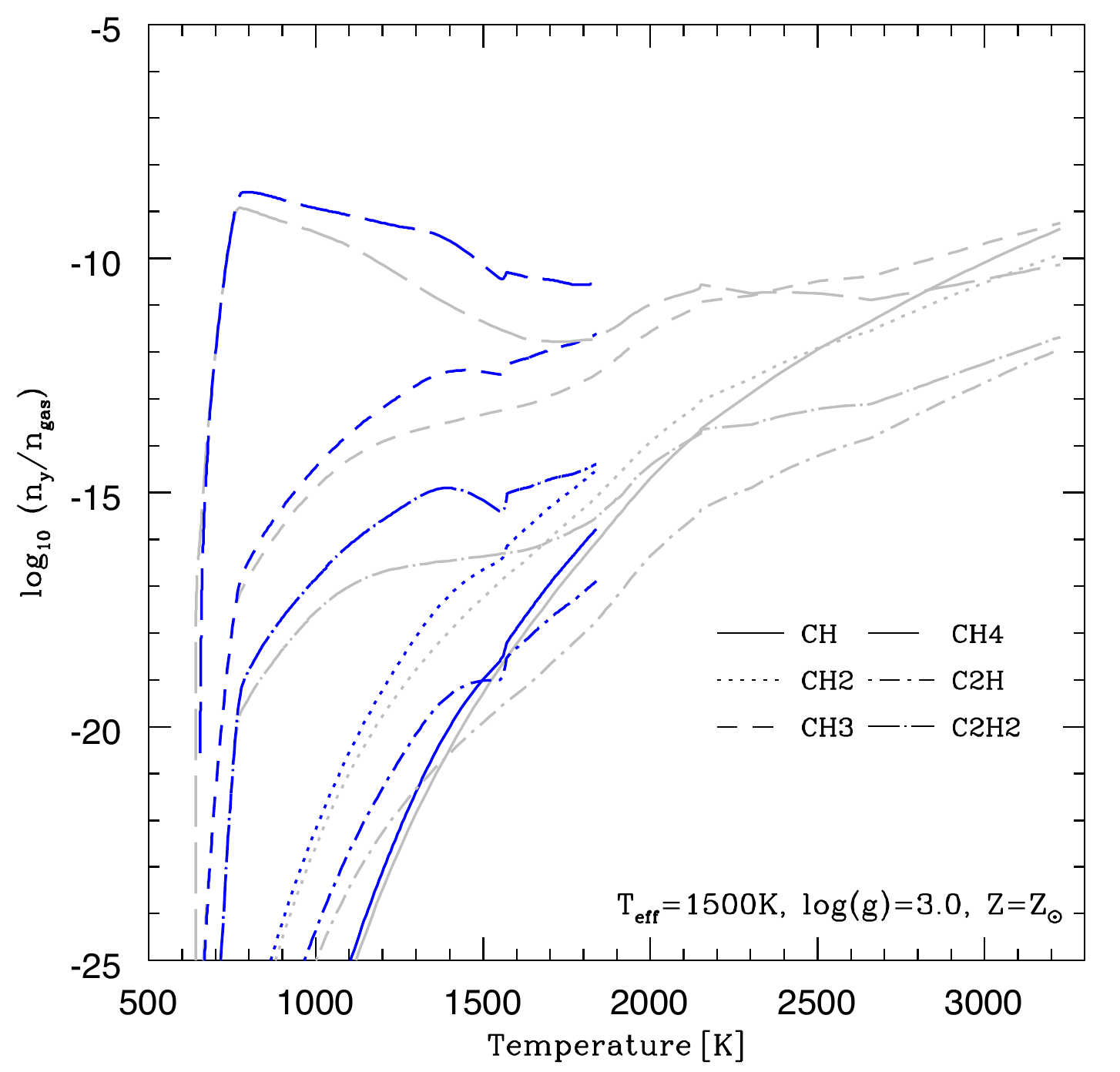}
\includegraphics[scale=0.39]{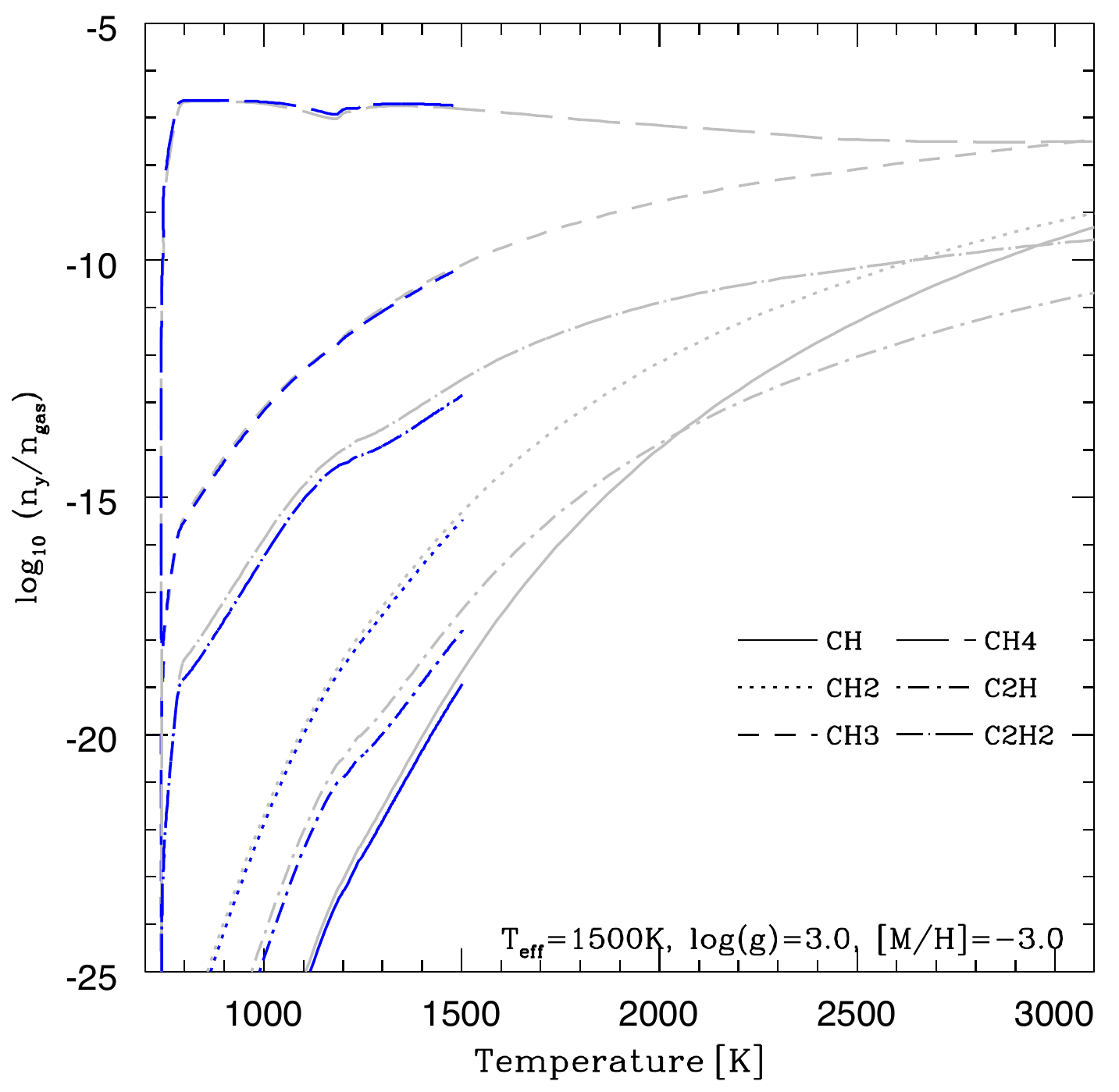}
\includegraphics[scale=0.39]{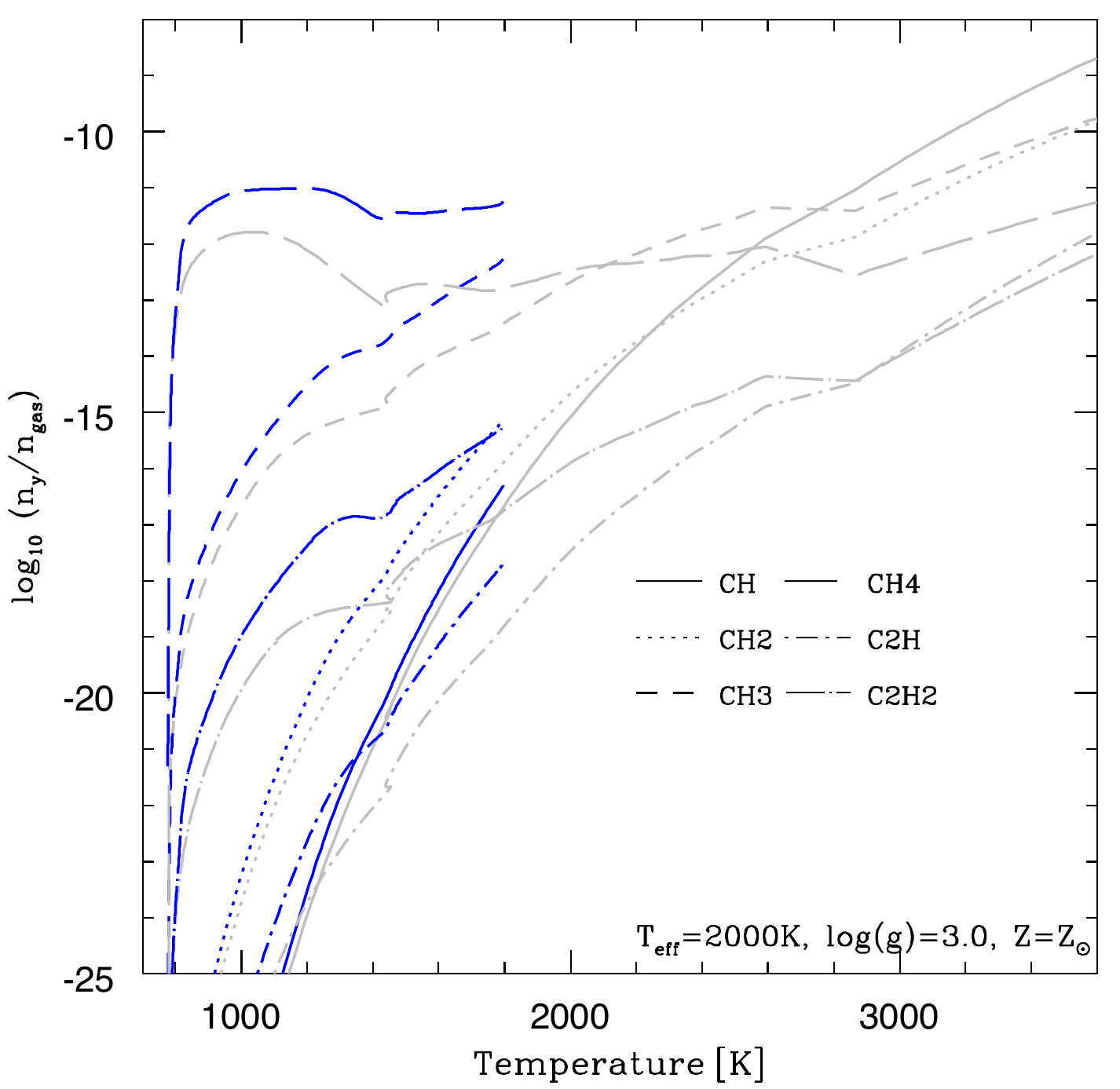}
\includegraphics[scale=0.39]{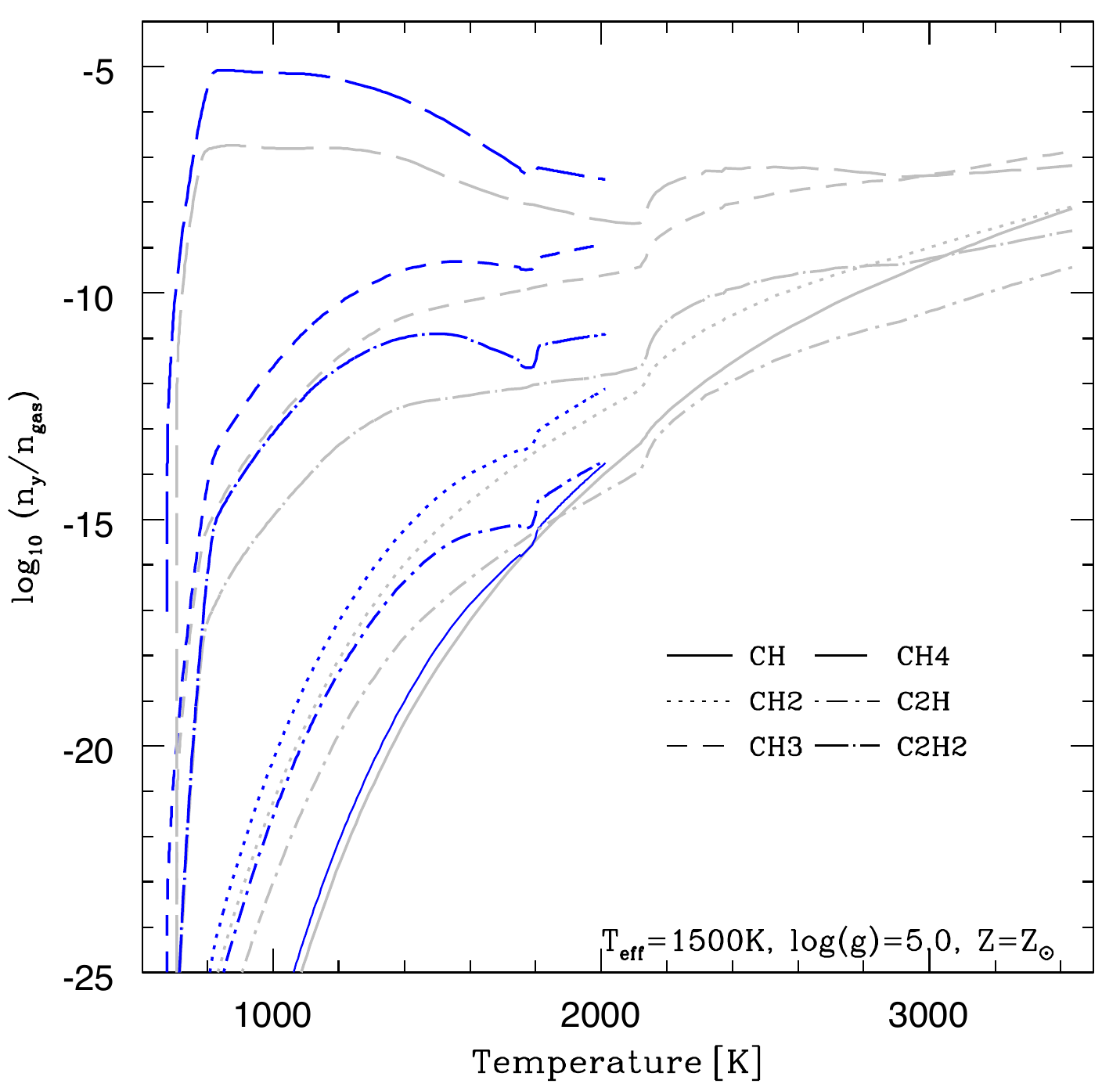}\\*[-0.3cm]
\caption{Gas composition for [C,H]-bearing gas-phase molecules: grey
  -- no element depletion (solar abundances), blue -- dust depleted
  element abundances. We utilise a giant gas planet {\sc
    Drift-Phoenix} model atmosphere (T$_{\rm eff}$=1500K, log(g)=3.0,
  initial solar). }
    \label{fig:GPc15003}
\end{center}
\end{figure}

\section{Chemical equilibrium results}
\label{sec:results}

 In what follows, we present our results of the composition of the gas
 of the deeper atmospheric layers, i.e. below the uppermost atmosphere
 which could be affected by photochemistry. We are interested in how
 cloud formation may indirectly impact the occurrence or the increase
 of the number density of complex carbon molecules in oxygen-rich
 environments. Such an indirect influence results from an
 inhomogeneous depletion of elements due to the formation of cloud
 particles, and from the feedback on the local temperature-pressure
 structure due to the large opacity of a dust cloud. We study the
 effect of the atmosphere model parameter (T$_{\rm eff}$, log(g),
 metallicity), and investigate a case for an artificially
 increased carbon over oxygen abundance.

We present the calculations of the gas-phase abundances for a given
atmospheric structure. The local temperature, pressure and element
abundances are a result of the {\sc Drift-Phoenix} atmospheric
simulations (see Sect. 2.1). In Sect.~\ref{ss:elmab} the local
gas-phase element abundances are discussed with respect to the element
depletion due to cloud formation. These element abundances are input
quantities for our chemical equilibrium routine.  Their impact on the
gas-phase compositions is demonstrated in Sect.~\ref{ss:refmod} for an
example of a giant-gas planet model atmosphere (T$_{\rm eff}=1500$K,
log(g)=3.0, initial solar element abundances). Section~\ref{ss:PAHs}
shows which carbon-binding macro-molecules, small PAHs and HCN
molecules could be expected in chemical equilibrium in the dense,
cloud forming part of a Brown Dwarf or giant gas planet's atmosphere.
Both differ by their gravitational surface acceleration (log(g)),
which causes the giant gas planet to have a much larger pressure scale
height in the atmosphere, hence, to be less dense than a Brown Dwarf
atmosphere. Section~\ref{ss:crich} tests how our results change if the C/O
ratio is increased to larger than one.


\subsection{Element abundances}
\label{ss:elmab}

The cloud formation as part of the {\sc Drift-Phoenix} atmosphere
simulations affects the metal elements O, Ti, Mg, Si, Fe, Al which are
depleted by the amount needed to form the cloud particles. This results
a metal-depleted gas-phase. This becomes apparent in
Fig.~\ref{fig:dustelem}: the metal elements decrease as dust forms.
Deeper inside the atmosphere, after evaporation of the dust grains,
the metal elements are released back into the gas-phase. The relative
depletion is independent on the initial element abundances as a
comparison between the upper (solar) and the lower (subsolar) panel of
Fig.~\ref{fig:dustelem} shows.

The depletion of the oxygen causes the carbon-to-oxygen ratio to
increase as relatively more carbon is available than without dust
formation.  Figure~\ref{fig:cto} demonstrates
further that the change in C/O from the solar value to $\sim 0.7$ is
independent of the stellar parameters T$_{\rm eff}$ and log(g). Only
the atmospheric gas temperature range concerned becomes wider with
increasing gravity, hence for Brown Dwarfs, slightly more oxygen
is available at the cloud base compared to a gas giant's atmosphere.

The drop below the initial value (solar or [M/H]$= -3.0$) of the C/O
ratio, is  caused by a rise in the number fraction of oxygen at the
cloud base following the gravitational settling and complete
evaporation of dust particles containing oxygen.

\begin{figure*}
\includegraphics[scale=0.55]{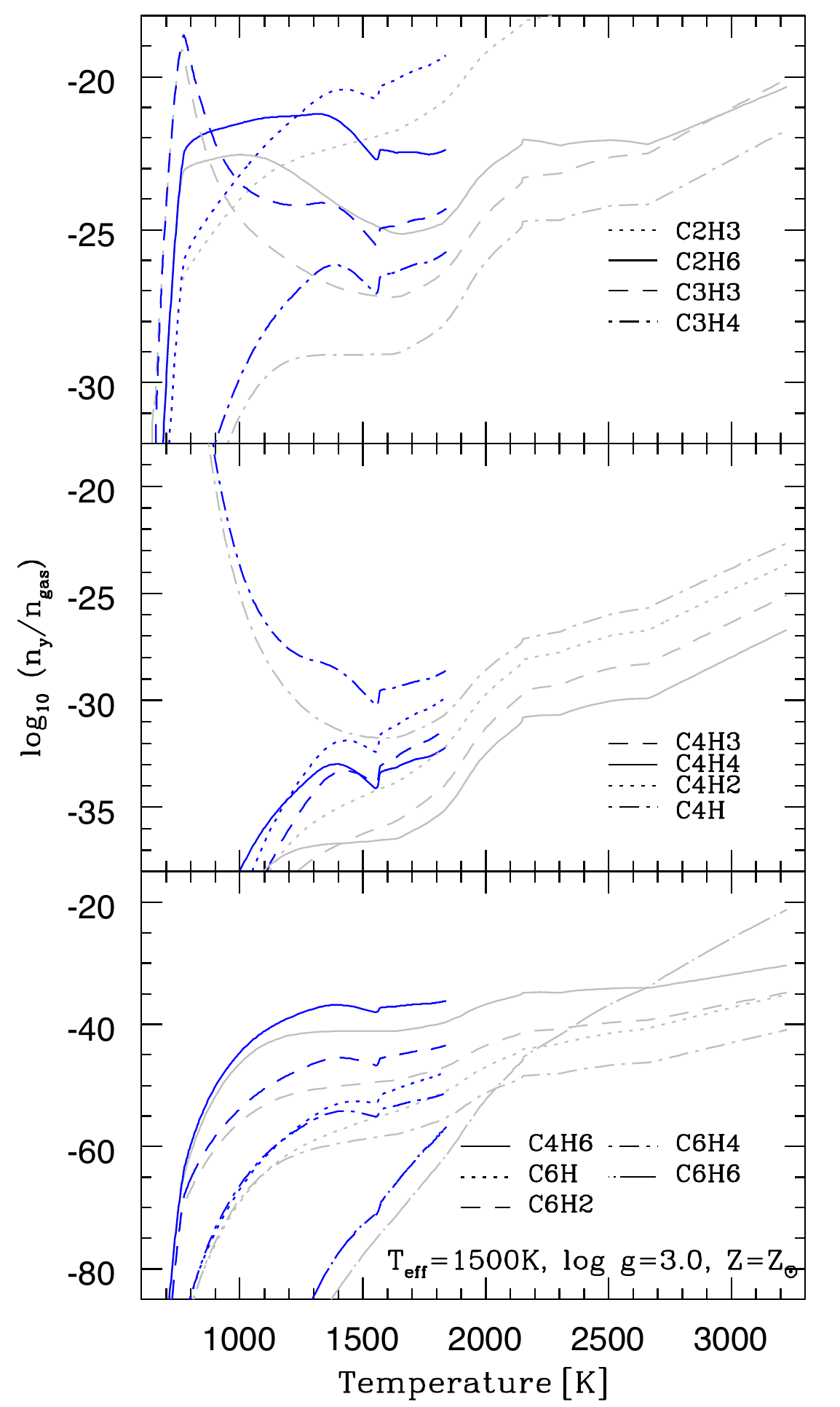}
\includegraphics[scale=0.55]{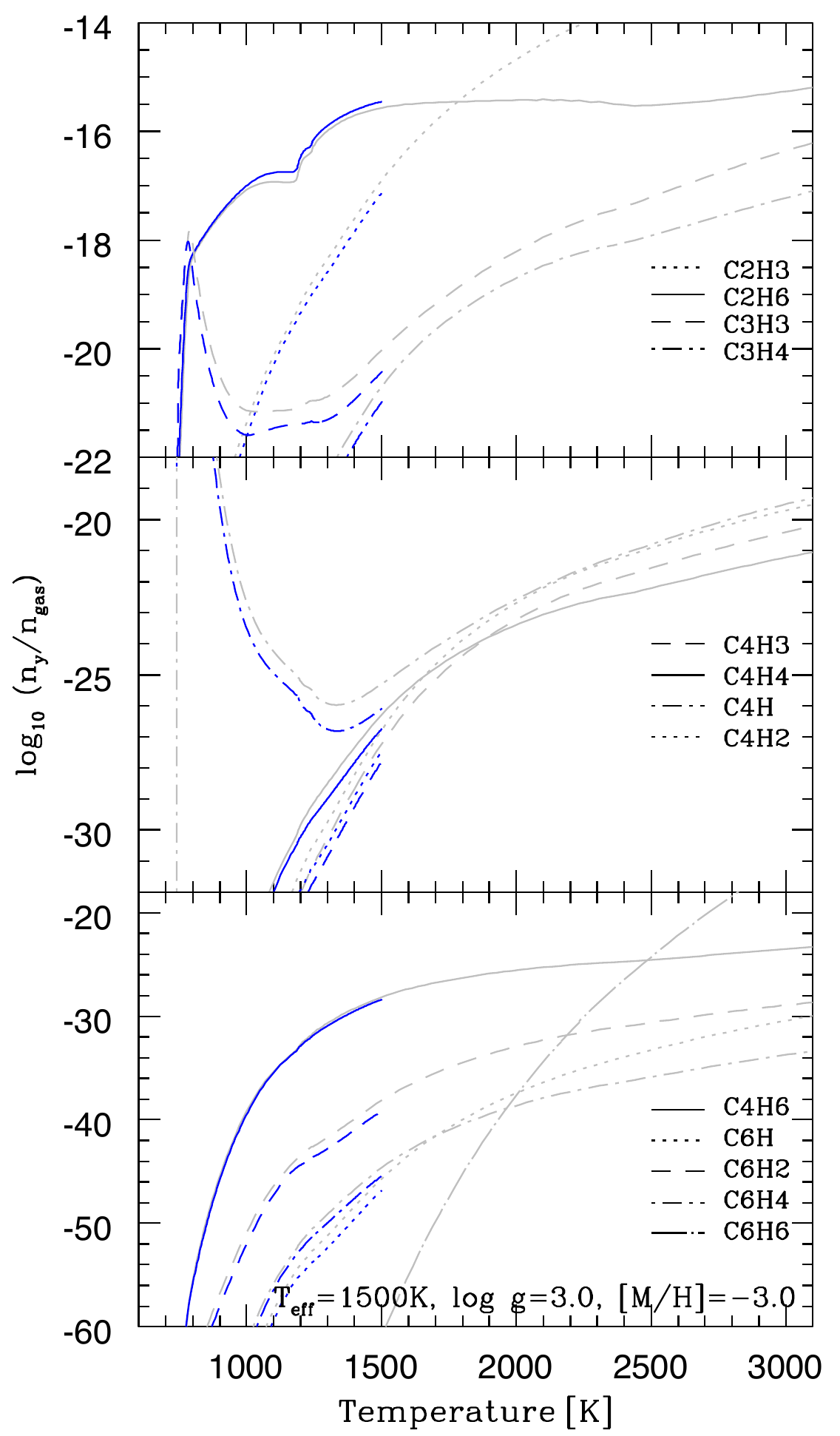}\\
\includegraphics[scale=0.55]{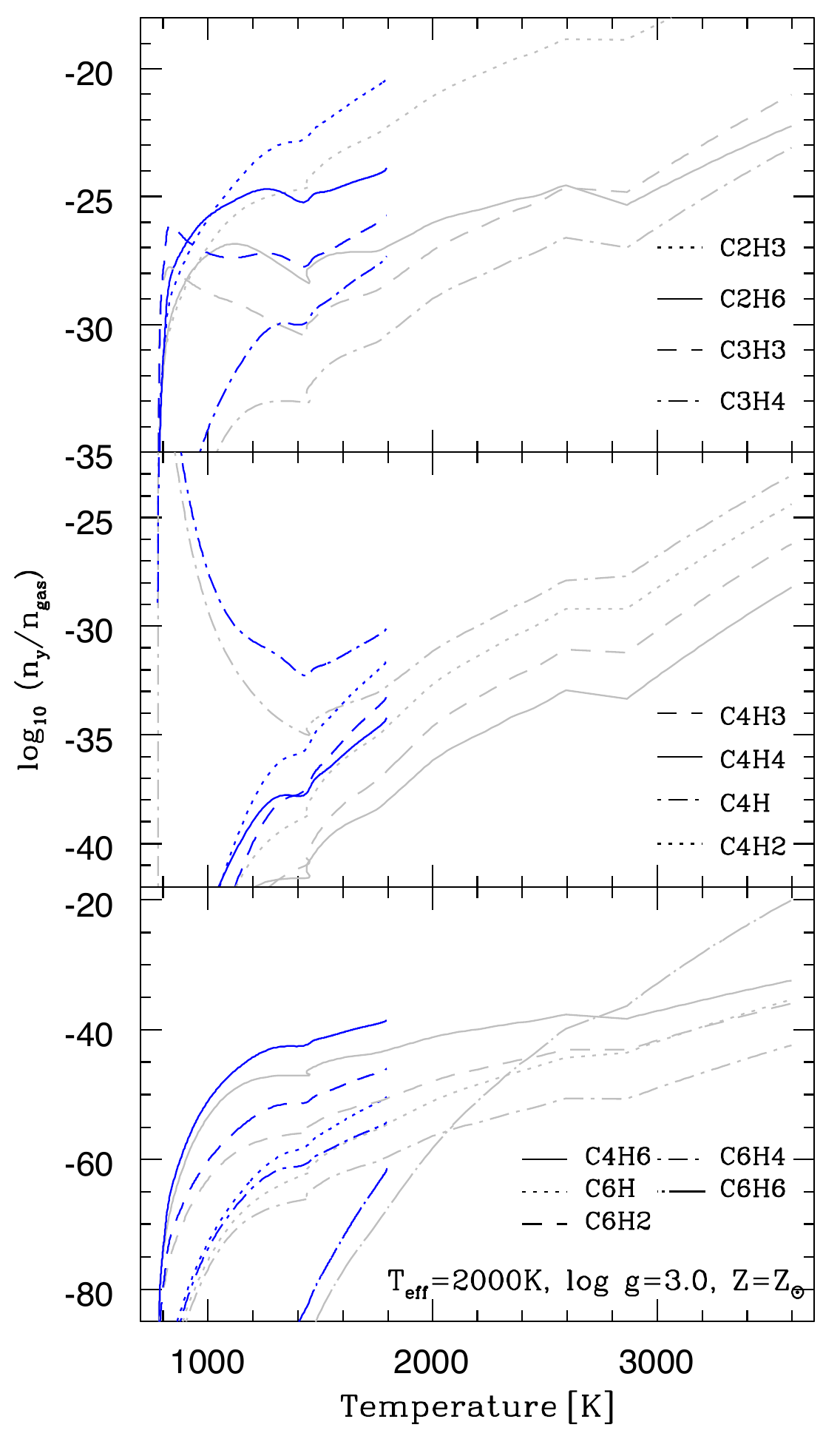}
\includegraphics[scale=0.55]{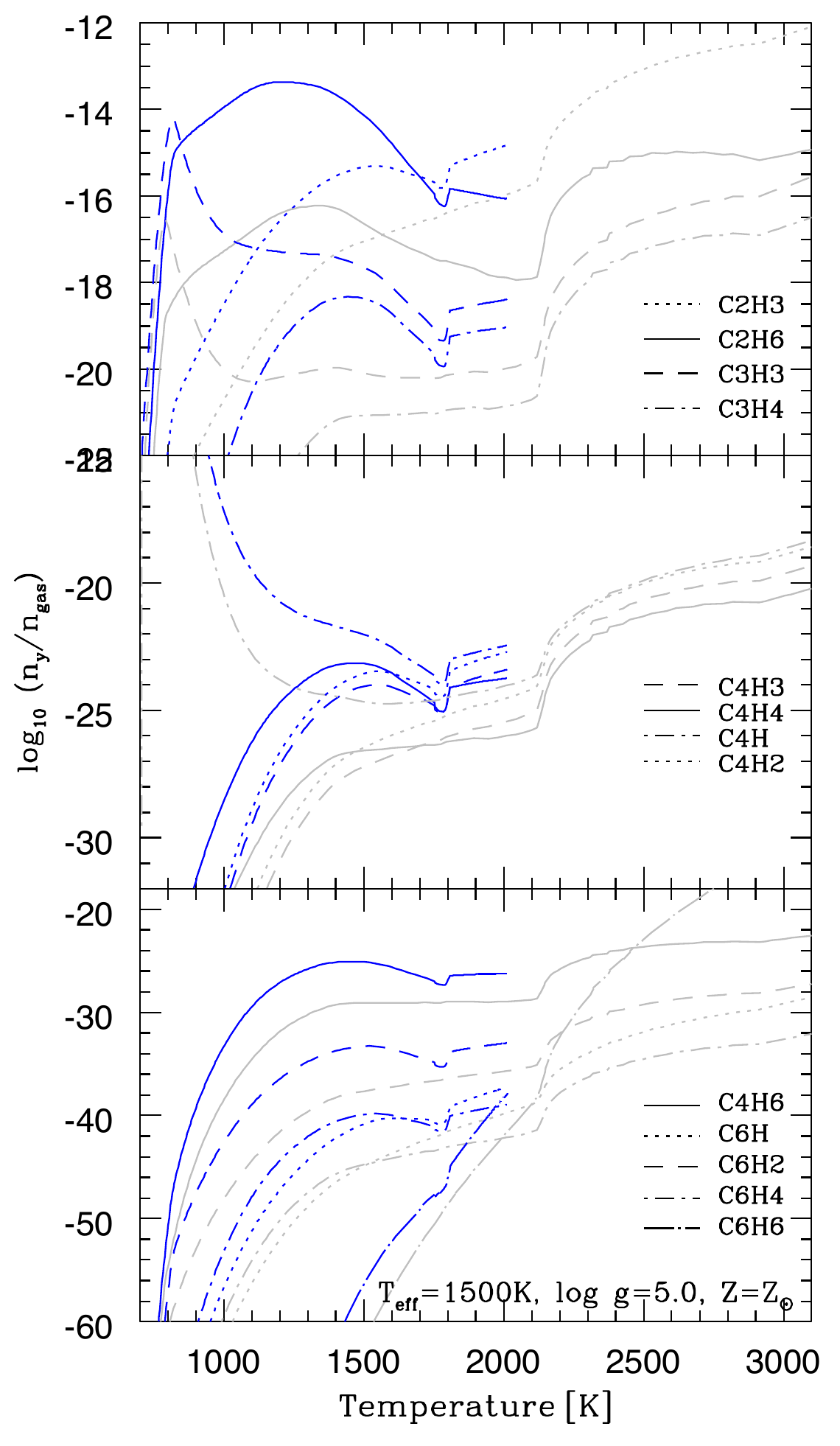}\\*[-0.3cm]
\caption{Same as Figure~\ref{fig:GPc15003} for small hydrocarbon molecules.
} 
   \label{fig:pahGP}
\end{figure*}

\subsection{Reference calculations}\label{ss:refmod}

The elemental abundances determine how much of a given element is
available to form molecules or ions. We have chosen one example model
to demonstrate how different the molecular abundances are for a
dust-depleted (blue) and a non-depleted (grey) element composition for
an oxygen-rich planetary atmosphere of T$_{\rm eff}=1500$K, log(g)=3.0
and (initial) solar metallicity (Figure~\ref{fig:others15003}).
Oxygen-bearing, dust forming molecules can be considerably less
abundant inside the cloud layers \sed{ which coincides with the
  deviation of the elements from their initial values, because the
  dust formation processes consumes the local elements. This is shown
  in Fig.~\ref{fig:dustelem} where the cloud layer spans a temperature
  range from T$_{\rm gas}=100\,\ldots\,\sim1600$K for the
  low-metallicity giant gas planet atmosphere (bottom panel).}

As the metal elements are depleted by dust formation, they become
unavailable to the formation of typical oxygen-rich gas-phase
molecules (e.g. SiO, FeO, MgOH). The Ti-bearing gases are the least
abundant; the Fe-, Al-, Si-, and Mg-bearing gases have concentrations
$\log(n_{\rm y}/n_{\rm gas}) \approx 10^{-5}\,\ldots\,\approx
10^{-15}$, hence they decrease by orders of magnitude inside the cloud
region. We observe further that molecules such as, e.g.  NH$_3$, which
are not directly involved in dust formation can change their
abundance if dust forms. This is largely due to the effect of dust
formation on the gas-phase metallicity, and its effect on
the atmospheric temperature structure.

Figure~\ref{fig:others15003} demonstrates the major reservoir
for oxygen in substellar oxygen-rich atmospheres are gaseous H$_2$O
and CO. CH$_4$ is the most abundant hydrocarbon (Fig.~\ref{fig:GPc15003}). Its concentration
($\approx 10^{-5}$-$10^{-10}$) is comparable to the concentration of
the typical oxygen-binding molecular species encountered in
substellar atmospheres.  We refer for a more complete plots
  regarding to oxygen-rich gas-phase abundances to e.g. \citet{Lodders2002}.


\begin{figure}
\centering
\includegraphics[scale=0.39]{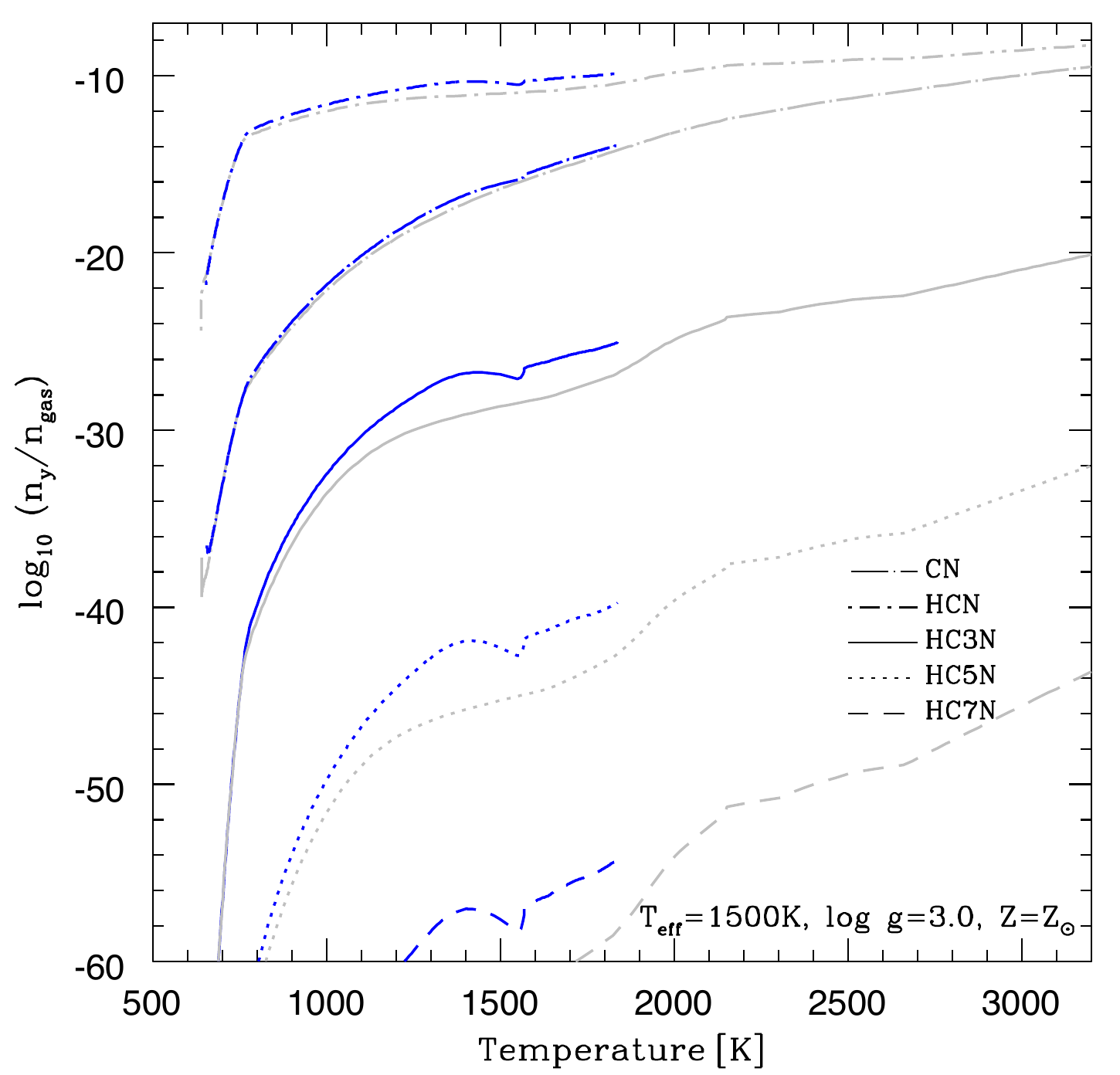}\\
\includegraphics[scale=0.39]{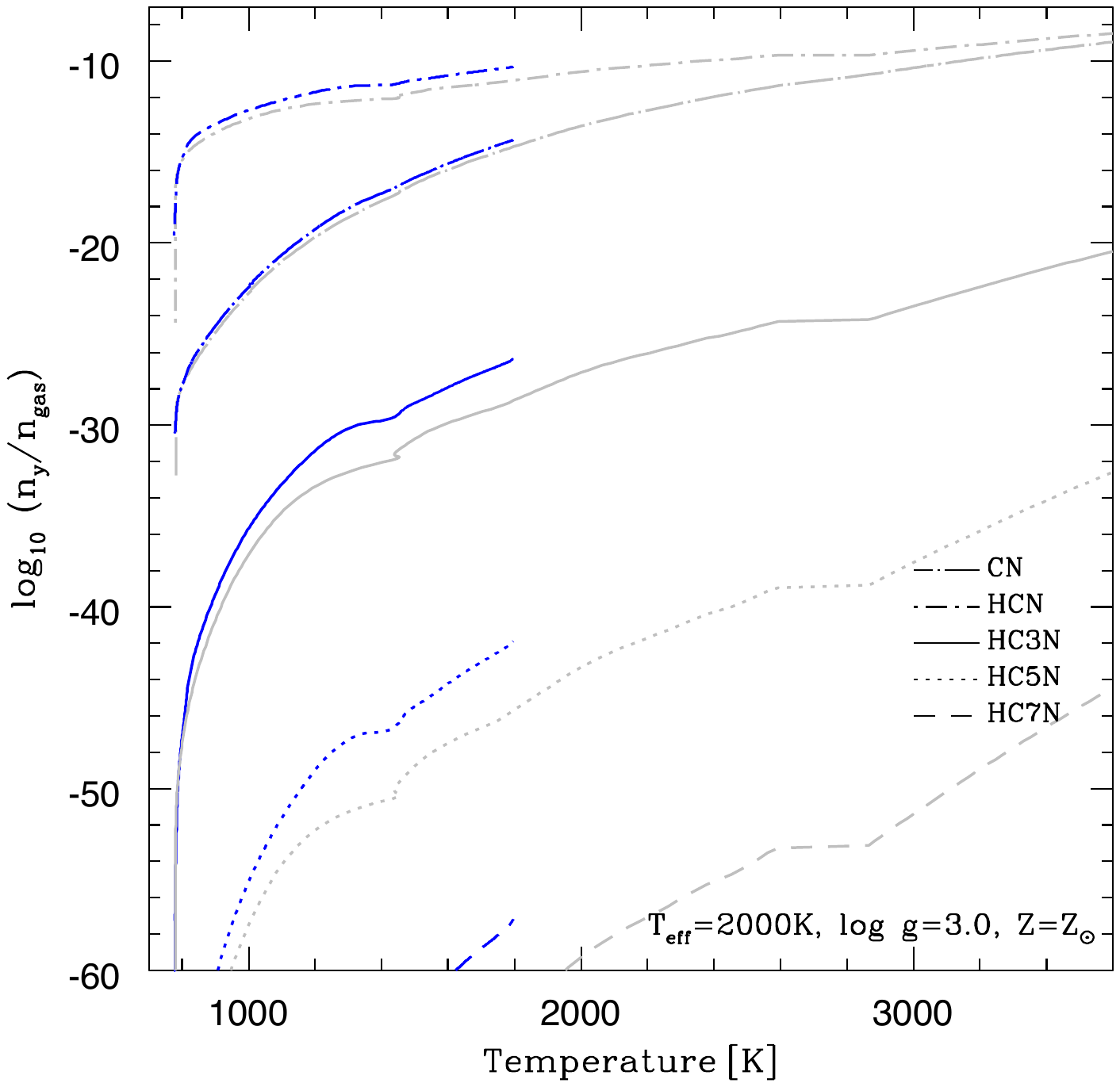}\\
\includegraphics[scale=0.39]{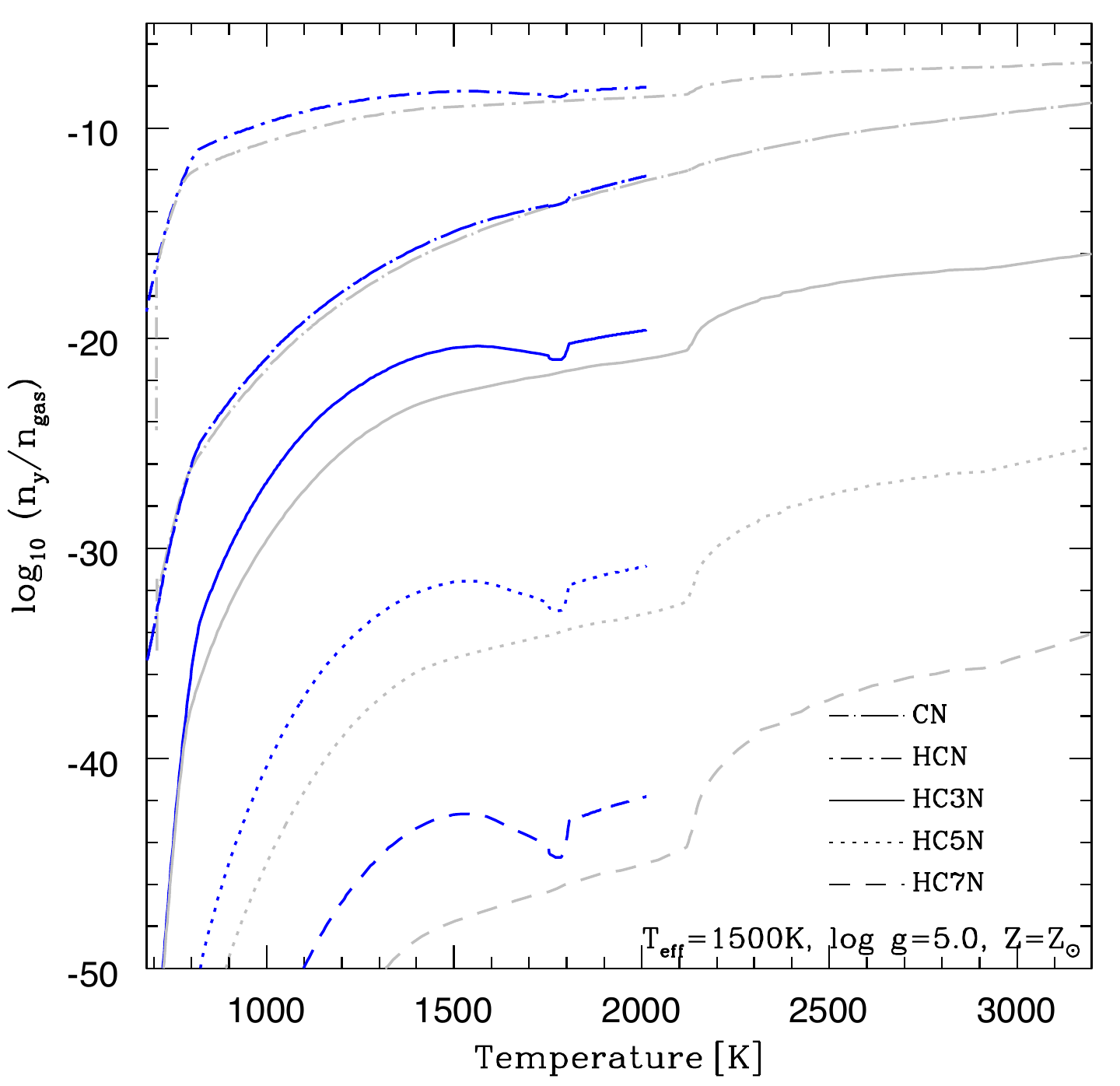}\\*[-0.1cm]
\caption{Molecular number densities, $\frac{n_y}{n_{\rm gas}}$, of typical [C,N]-bearing gas-phase 
molecules for a giant gas planet and a Brown Dwarf based on {\sc Drift-Phoenix} model atmospheres.
The HC$_x$N are generally low abundant and not much changes amongst the models investigated in this paper.}
    \label{fig:hcn1500}
\end{figure}

\subsection{Hydrocarbons \& PAH's}\label{ss:PAHs}


Carbon and hydrocarbon macro-molecules are not included in the dust
cloud formation considered in the {\sc Drift-Phoenix} model
atmospheres, nor in any of the other model atmosphere simulations for
substellar objects. But dust formation has a strong impact on the
oxygen abundance. Because of the large abundance of oxygen compared to
iron, magnesium etc, the change seems small at a first glance, but the
resulting C/O ratio is considerable
(Sect.~\ref{ss:elmab}).  The question is: how many carbon-bearing
macromolecules would we expect in the collisionally dominated
chemistry of a denser atmospheric environment, and how would this
change if an oxygen-depleting process like cloud formation occurred?  
Our calculations will serve as an inner boundary for
more complex, kinetic models that are not yet able to resolve all
possible reaction paths, and that address the outer atmospheric
regions only (Zahnle et al. 2009, Line et al. 2010, Moses et al. 2011,
Venot et al. 2012).


We find that inside the cloud layer where oxygen is depleted, the
carbon-bearing molecules, including the CN-complex, are more abundant
than outside the cloud layer. For example, C$_2$H$_6$ increases by $3$
orders of magnitude in concentration, C$_4$H$_4$ by $4$ orders of
magnitude and CH$_4$ by $1.5$ order of magnitude
(Figs. \ref{fig:GPc15003}, \ref{fig:pahGP}; T$_{\rm eff}$=1500K,
log(g)=3.0 and solar). CO, on the other hand, does not vary
visibly (Fig.~\ref{fig:others15003}). The effect is smaller for the
  CN-molecules (Fig.~\ref{fig:hcn1500}).

All number densities rise, with the exception of C$_4$H, with local
gas temperature as the gas density increase inside the atmosphere. A
higher density allows more molecules to form as the atmospheric
number density increases. The smaller member of the hydrocarbon groups is
always the most abundant. However, C$_6$H$_6$ (the first
\emph{aromatic ring}, or PAH) becomes one of the most abundant
hydrocarbons in the deeper atmospheric layers. Its number density is
comparable to the one of TiO or to hydrocarbons bearing 2 or 3 carbon
atoms (Fig.~\ref{fig:others15003}).  The results may appear
  surprising, but the atmospheric environments that we
  investigate here are considerably more dense than for example the
  atmosphere of an AGB star by which the study of carbon-molecules was inspired.


The PAHs, other than C$_6$H$_6$, do not form in significant
quantities.  The little ``dip'' observed in each curve is due to the drop
of the C/O ratio at $\sim 1550$K (model dependent). It is attributed
to a sudden small rise in the number fraction of oxygen upon elemental
oxygen replenishment of the cloud base following the gravitational
settling and complete evaporation of dust particles containing oxygen.

\begin{figure*}
\begin{center}
 \includegraphics[scale=0.46]{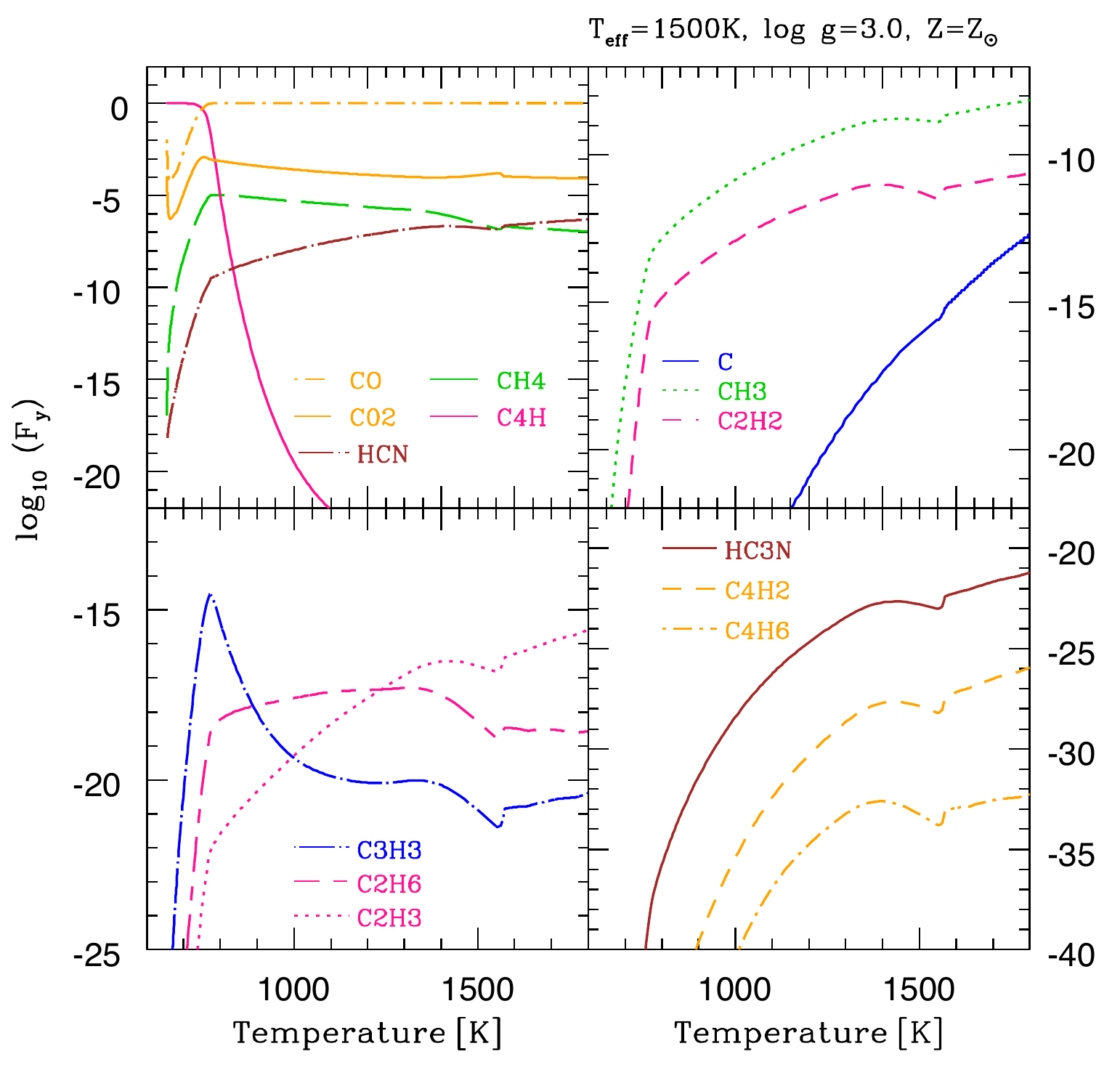}
\includegraphics[scale=0.46]{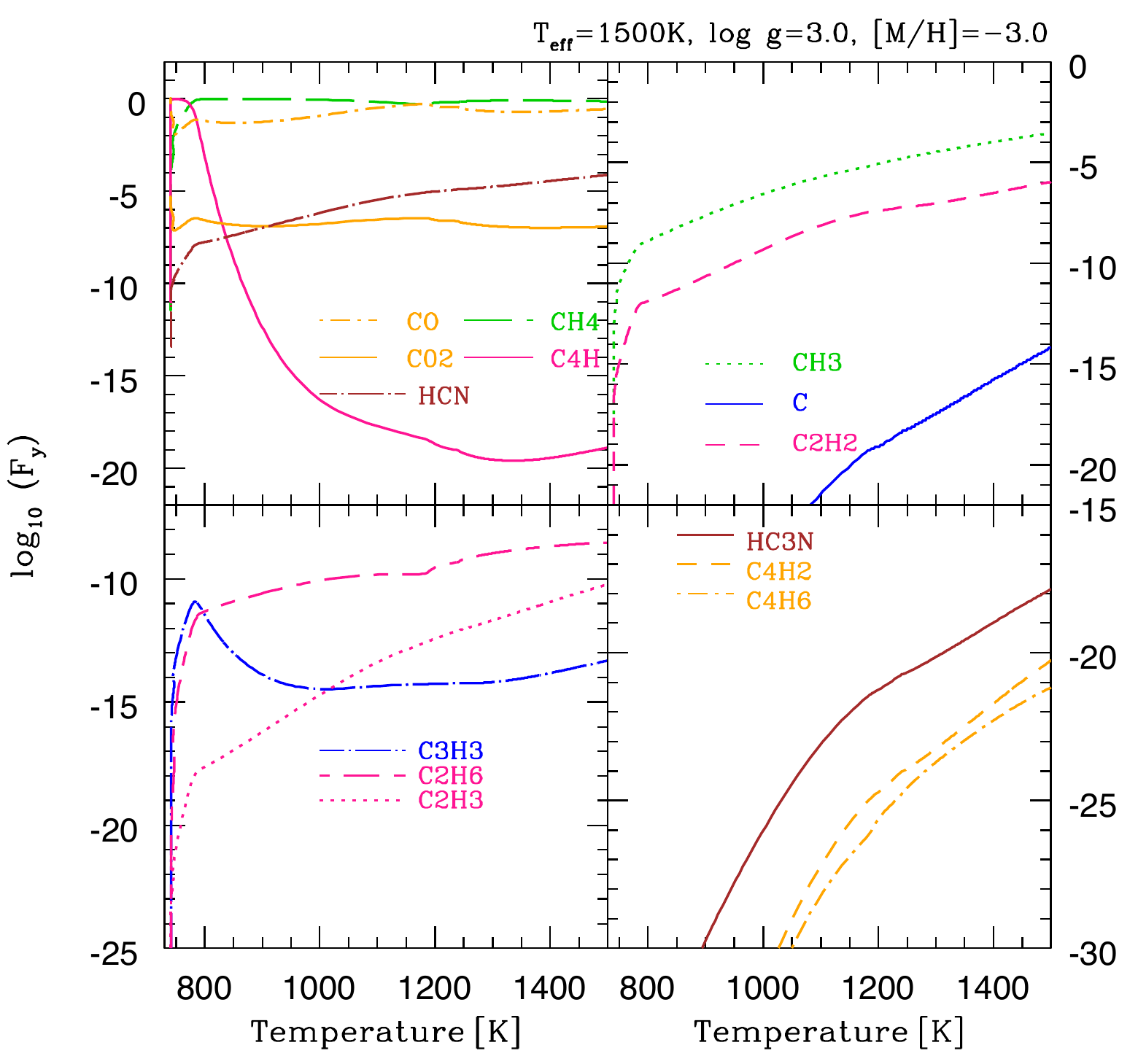}\\
\includegraphics[scale=0.46]{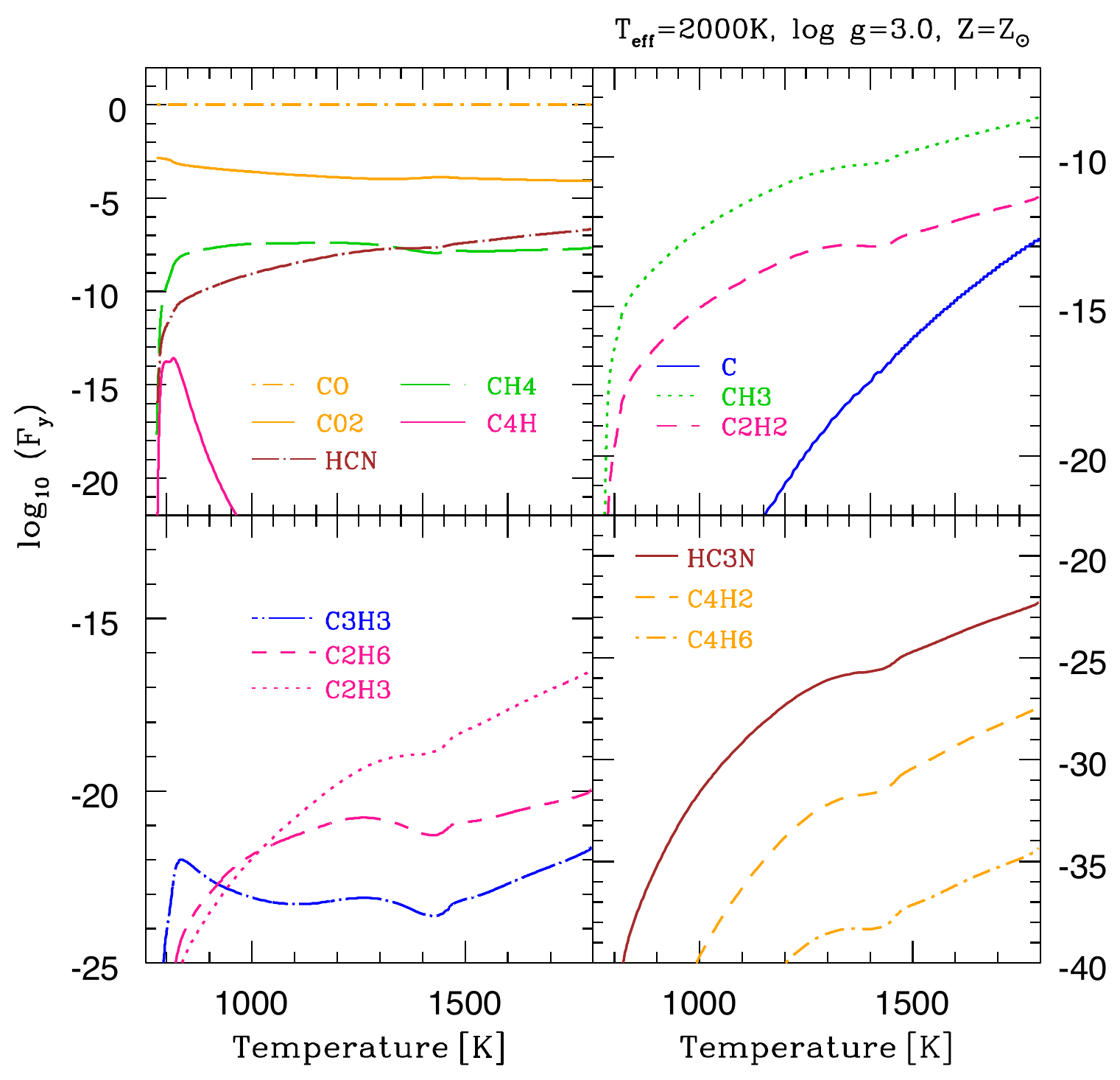}
\includegraphics[scale=0.46]{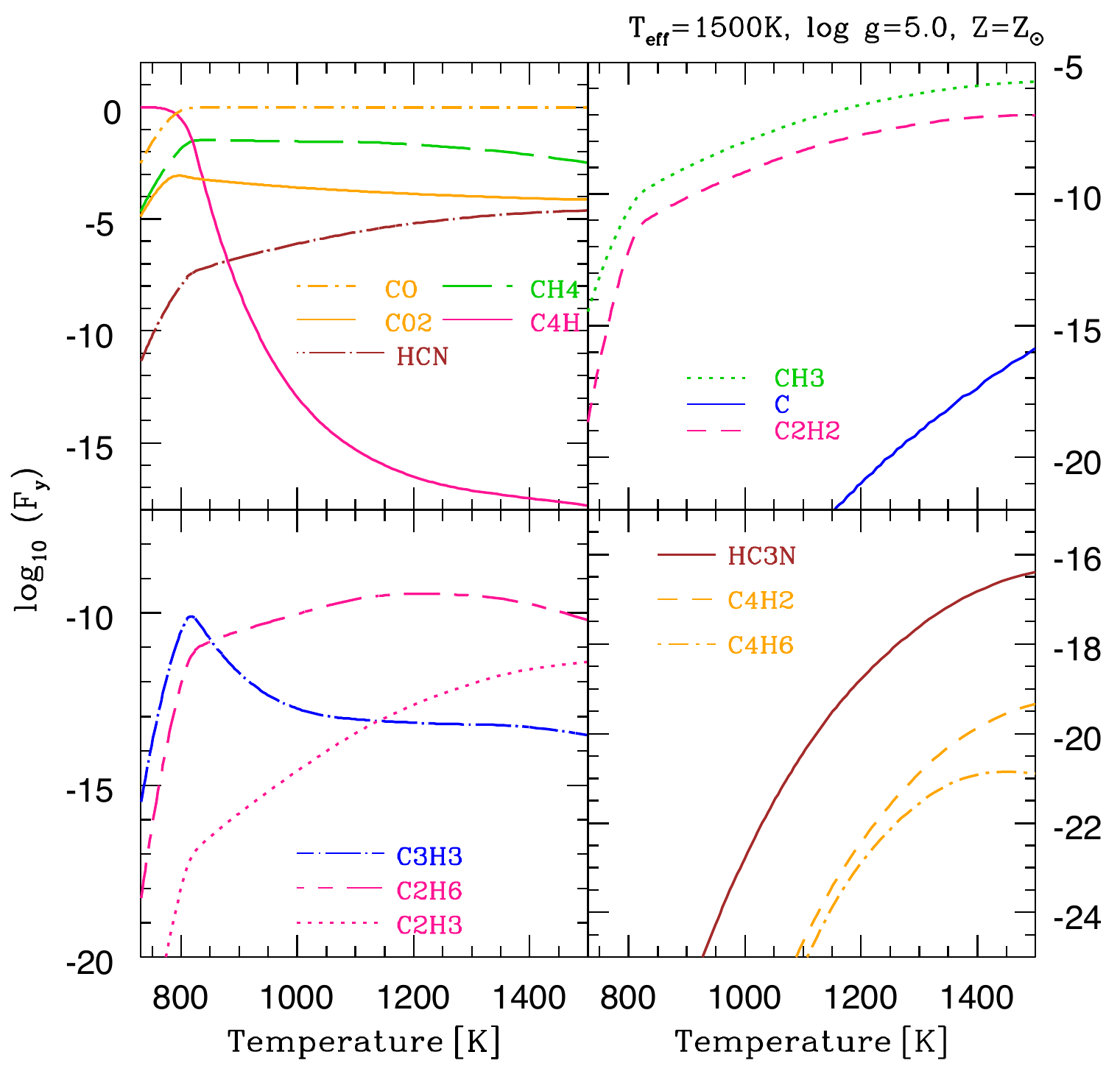}\\*[-0.1cm]
\end{center}
\caption{Fraction, $F_y$, of carbon in species $y$ for the cloud layer
  of a giant gas planet atmospheric model of solar composition ($\rm
  T_{eff}=1500 \rm K, 2000 \rm K$) and low-metallicity composition
  ($\rm T_{eff}=1500\rm K$) based on a {\sc Drift-phoenix}. Below $\rm
  T \sim 500$K, more carbon is bound in the hydrocarbon species C$_4$H
  than in CO.}
    \label{fig:fracGP}
\end{figure*}

\textbf{Varying T$_{\rm eff}$, log(g) and [M/H]:} We study the
influence of T$_{\rm eff}$, log(g) and [M/H] on the molecular
concentrations. An increase of the effective temperature to T$_{\rm
  eff}=2000$K causes decreasing number densities of the
macro-molecules considered here. This is a consequence of a lower gas
pressure for a given temperature in the atmosphere.
 It must be noted
that all carbon-bearing molecular abundances remain greater inside the
cloud layers  than outside, even for increasing T$_{\rm eff}$.

Carbon-bearing species are able to form in higher abundances inside
the cloud layers for an increasing surface gravity, i.e. inside Brown
Dwarfs' atmospheres. This is a consequence of the higher gas pressure
throughout their atmosphere compared to a gas giant's
atmosphere. Increasing the surface gravity increases the partial
pressures of the gas-phase molecules and hence their number
densities. The pronounced jump in the number densities at $\sim 2150$K
coincides with the (T,~$p$)-structure of the model. It is a feedback
that results from dust formation causing a backwarming effect.
In spite of this increase in gas pressure, hydrocarbons with $6$ or
$4$ carbon atoms are still very rare ($< 10^{-20}$ in
concentration). Benzene is again an exception in the deeper
atmospheric layers, where it reaches a concentration of $10^{-9}$ \sed{
  for the very low-metallicity cases}, which is comparable to the
oxygen-bearing molecules' concentration shown in
Fig.~\ref{fig:others15003}. \sed{ Note, that low-metallicity
  atmospheres are considerably denser that solar metallicity
  atmospheres as demonstrated by Fig.~\ref{fig:TP}.}

The metallicity is another parameter that determines the structure of a
(model) atmosphere, but it is not very well constrained as our results in
Sect.~\ref{ss:elmab} suggest. The metallicity parameter is often
introduced because detailed knowledge of individually varying element
abundances is only very scarcely available. Works by Burgasser and
co-worker
show that metallicity needs to be considered as a
parameter also for Brown Dwarfs and giant gas planets.

 By decreasing the metallicity, [M/H], atomic hydrogen becomes even
 more abundant relative to heavier elements. The concentrations of the
 different H-bearing species throughout the atmosphere increase
 compared to the solar composition case.  Molecules with a higher
 number of H-atoms (e.g. CH$_4$ and C$_2$H$_6$) remain in higher
 concentration inside the cloud layer, although by a much smaller
 fraction than in the solar-metallicity models studied. This small
 surplus in the concentration of some hydrocarbons leads to a lowering
 of the abundances of the other carbon-bearing molecules and
 compensates for the disappearance of \sed{ hydrogen-saturated carbon-containing species}.

\subsection{The carbon fraction locked up in carbonaceous molecules}

In substellar atmospheres, most of the hydrogen is locked in H$_2$,
and most of the carbon is in CO or CH$_4$. In this section, we
calculate the carbon fraction locked in some hydrocarbon and
cyanopolyyne molecules to test for the presence of alternative dominant
equilibrium forms of carbon under the temperature and pressure
conditions considered in this work following the work by Helling et
al. (1996). To calculate the fraction of carbon, by mass, locked into
each carbon-bearing molecule, we used
\begin{equation}
F_y = \frac{N_C \times m_H \times [y]}{\rho_{C}}
\end{equation}
where $N_C$ is the number of carbon atoms in molecule $y$, $m_{\rm H}$
is the mass in grams of a carbon atom, $[y]$ is the number density of
molecule $y$ \sed{ [cm$^{-3}$]} and $\rho_C$ is the total carbon mass density \sed{ [g\,cm$^{-3}$]} in the
gas-phase.

Figure~\ref{fig:fracGP} depicts the fraction F$_y$ of carbon atoms
bound in species $y$ along the temperature profile of the cloud layer
of a giant gas planet with T$_{\rm eff}$=1500K, log(g)=3.0 and solar
element abundances.  It is remarkable that, in the outer layers, C$_4$H traps most of the
available atomic carbon, before dropping exponentially. From $500$K
onwards, CO, CO$_2$ and CH$_4$ establish themselves as the dominant
carbon-bearing molecules throughout the cloud layer.

\textbf{Influence of T$_{\rm eff}$, log(g) and [M/H]:} When the effective
temperature is increased to $2000$K \sed{ (lower left panel in Fig.~\ref{fig:fracGP})}, F$_y$ generally diminishes since
the molecular number densities have decreased. F$_{\rm C_4H}$
substantially decreases and substitutes CO and CO$_2$ as the main
carbon-bearing species in the very outer layers of the atmosphere. In
a Brown Dwarf (log(g)$=5.0$), CH$_4$ is substituted as a more
important carbon-bearing molecule than CO$_2$.

\noindent In a low metallicity environment \sed{ ([M/H]=-3.0)}, CO and CO$_2$ are
overtaken by CH$_4$. This effect is more pronounced in high surface
gravity objects. Due to the favourable conditions to the formation of
H-bearing molecules, C$_2$H$_6$ is now a more important atmospheric
reservoir for carbon than C$_2$H$_3$ and C$_3$H$_3$ throughout the
entire cloud layer.

\begin{figure}
\begin{center}
\includegraphics[scale=0.39]{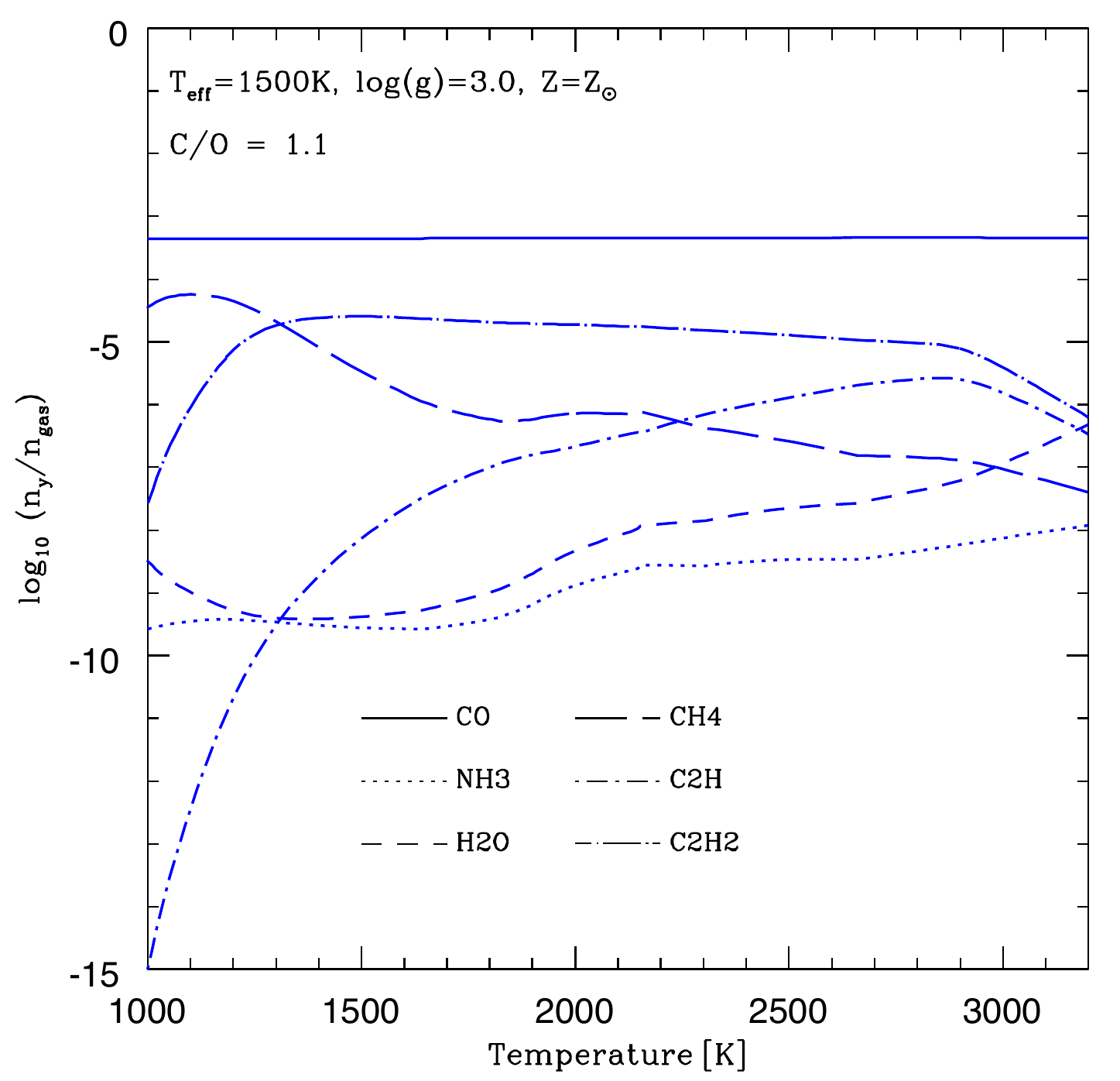}
\includegraphics[scale=0.39]{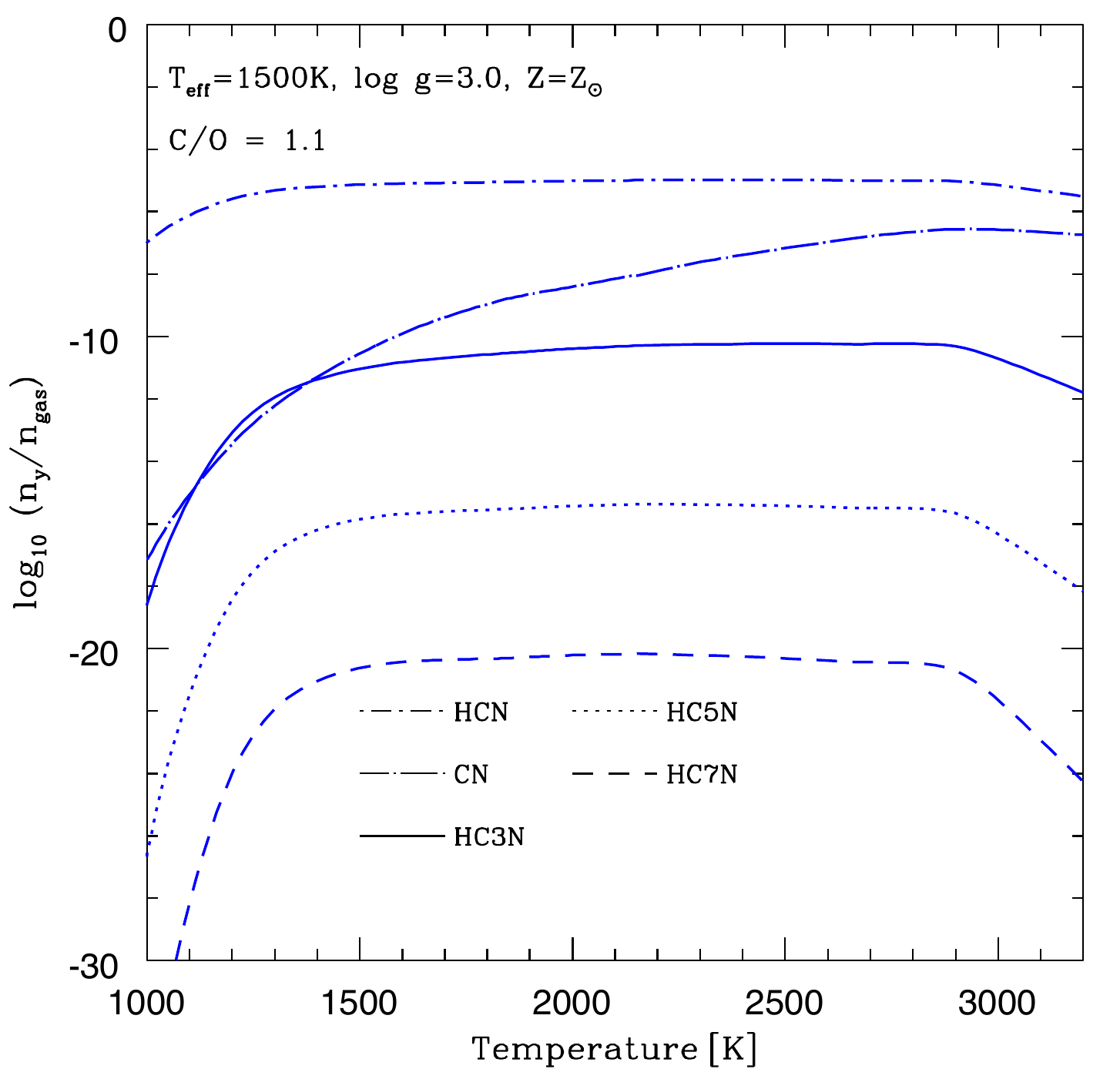}
\includegraphics[scale=0.39]{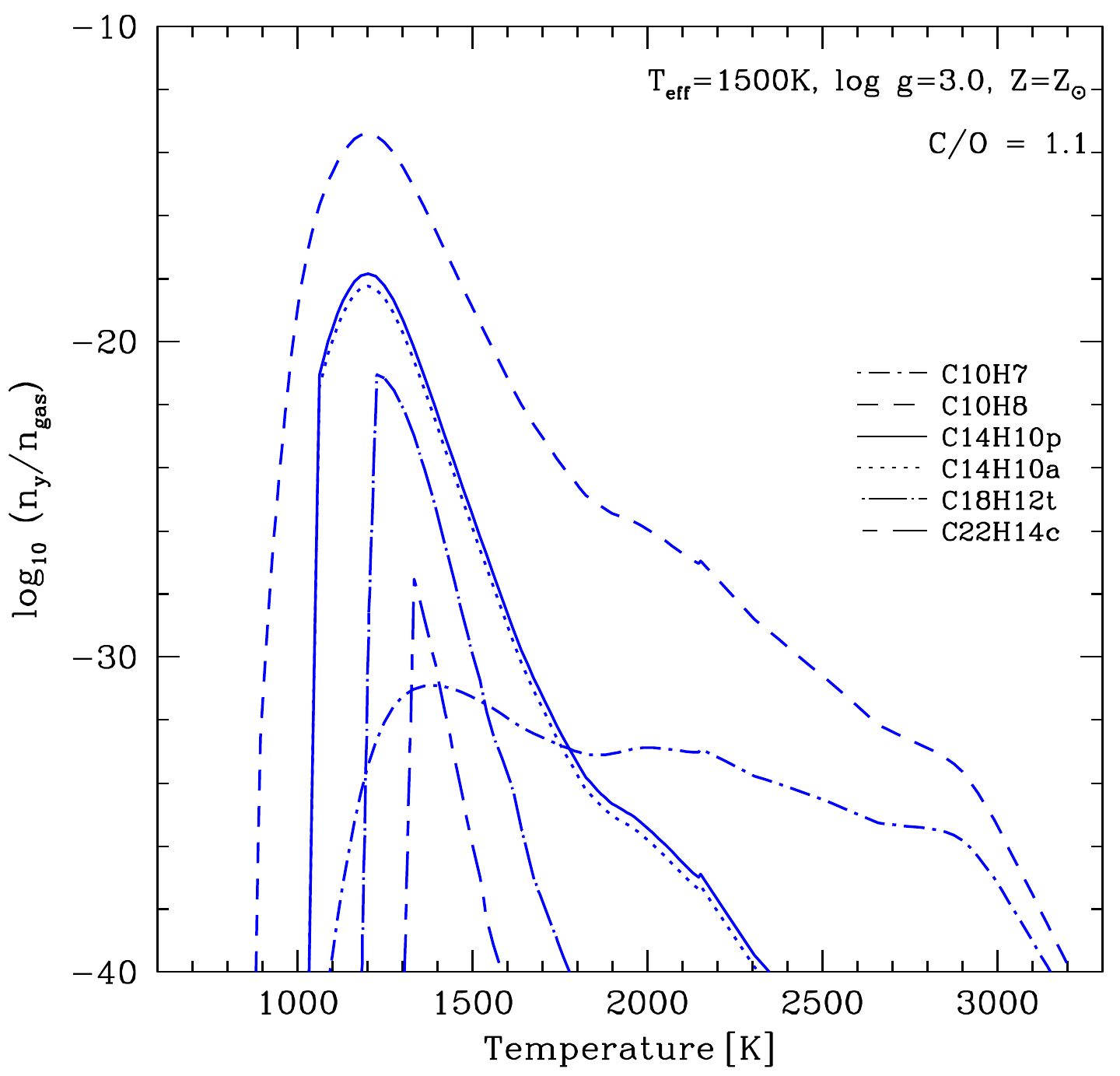}\\*[-0.1cm]
\end{center}
\caption{Chemical equilibrium abundances for small and large
  carbon-bearing molecules for a carbon-rich (C/O= 1.1)
  giant gas planet based on a {\sc Drift-Phoenix} model atmosphere for
  T$_{\rm eff}$=1500K, log(g)=3.0 and solar element abundances.}
\label{fig:co1.1first}
\end{figure}

\begin{figure}
\begin{center}
\includegraphics[scale=0.55]{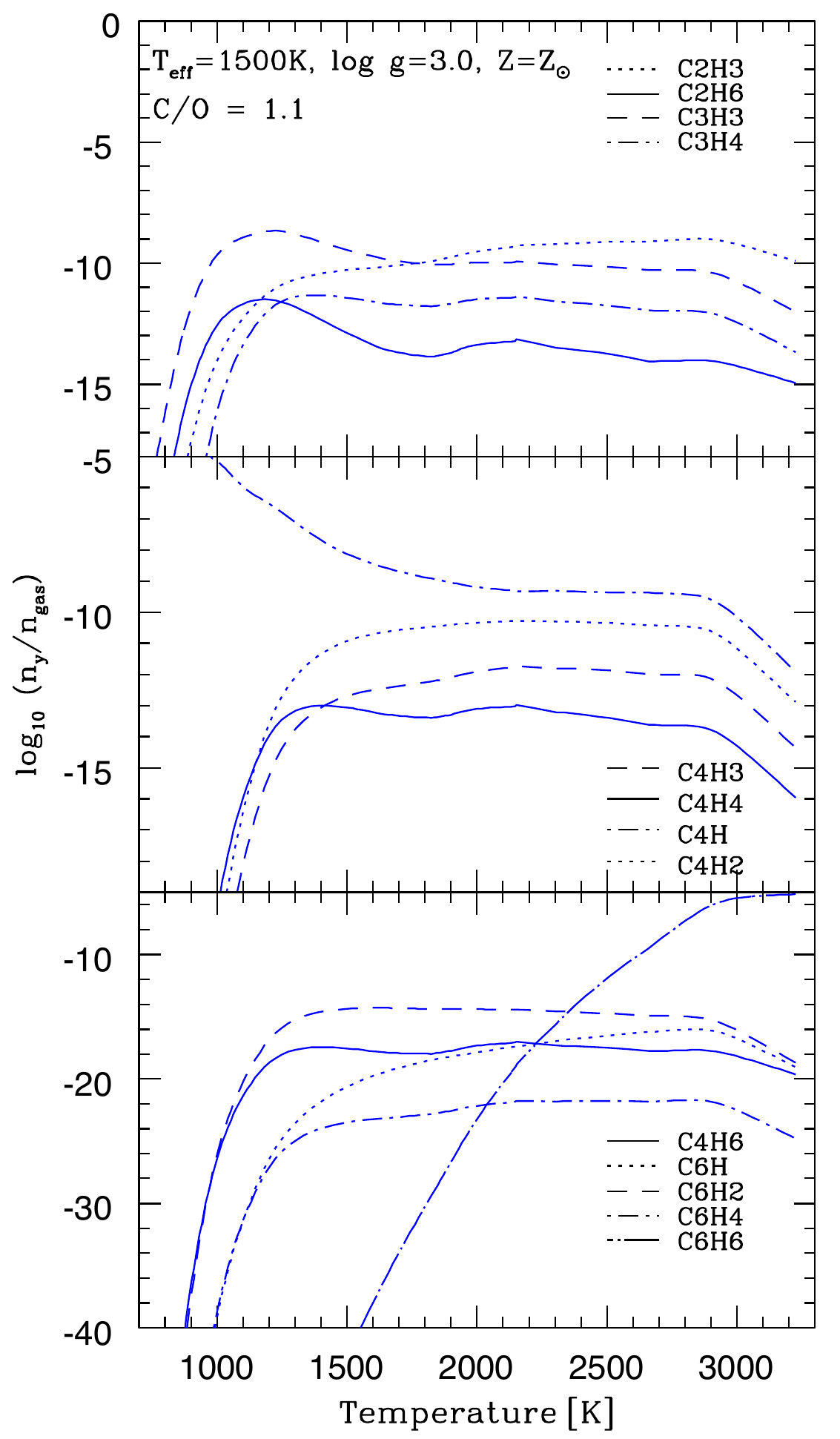}
\end{center}
\caption{Hydrocarbon abundances for a {\sc Drift-Phoenix} model
  gas-giant atmosphere of T$_{\rm eff}$=1500K, log(g)=3.0, solar element
  abundance sand also artificially increased carbon-abundances, such that
  C/O=1.1.}
\label{fig:co1.1second}
\end{figure}

\subsection{Carbon-rich substellar atmospheres}\label{ss:crich}

In extrapolation of our results so far, we show how the gas-phase
composition would change if more carbon becomes available in an
example giant gas planet atmosphere.  We use the same {\sc
  Drift-Phoenix} model atmosphere structures as before (T$_{\rm
  eff}$=1500K, log(g)=3.0 and solar element abundances), but now we
change the carbon element abundance such that more carbon than oxygen
is available, hence C/O=1.1 in our equilibrium chemistry routine
(Figure~\ref{fig:co1.1first}). No feedback onto the radiative transfer
is taken into account.

In comparison to an oxygen-rich atmosphere (C/O$<1$), the hydrocarbons
and the PAHs are significantly more abundant if C/O$>1$ even only
moderately. For example, the number densities of CN and C$_2$H$_2$
have increased significantly by 5 and 10 orders of magnitude,
respectively (compare Sect.~\ref{ss:PAHs}). C$_2$H$_2$ is now as
abundant as the typical molecules encountered in oxygen-rich
environments (compare Fig.~\ref{fig:others15003}).  C$_2$H$_2$ now
dominates over H$_2$O$\;$ throughout the entire atmosphere while CH$_4$
dominates over H$_2$O$\;$ until T=3000K. Figure~\ref{fig:co1.1first}
shows further that carbon-monoxide is the most dominant C-bearing
species in the model atmosphere used for this study. H$_2$O and NH$_3$
are of comparable abundance.

The importance of the larger amongst the PAHs has increased
  considerably by increasing the carbon such that C/O$>$1 (lower
  panel, Fig.~\ref{fig:co1.1first} and Fig.~\ref{fig:co1.1second}).

\citet{Helling1996} predicted the formation of PAHs with large
concentrations in the layers of dynamical carbon-rich stellar
atmospheres when T $\leq$ 850K (for higher effective temperature and
smaller surface gravity than the models studied in the present
work). In our study, the PAHs are important between a gas temperature
of 1000K and 1500K (Fig.~\ref{fig:co1.1first}). We recovered
the same molecular concentration for C$_6$H$_2$, C$_2$H$_2$ and HCN as
\citet{Helling1996} throughout the atmosphere; whereas CH$_4$ and
HC$_3$N are higher in concentration in our model giant gas planet.


\section{Departure from chemical equilibrium by convective mixing}
\label{sec:mixing-results}

So far, we have studied how abundant carbon-binding macro-molecules
can be in local chemical equilibrium in brown dwarf and giant gas
planet atmospheres.  Our results suggest that C$_2$H$_6$ becomes
gradually more important than C$_2$H$_2$ in low-metallicity
atmospheres as relatively less carbon is available.  We now study
possible deviations from the thermochemical equilibrium values by a
simple time-scale comparison.

  The major carbon-binding molecules in a solar metallicity gas, CO and
  CH$_4$, are rather small. Both, however, have been shown to be
  affected when vertical mixing processes are faster than their
  destruction kinetics. According to \citet{Moses2011} and
  \citet{Prinn1977} transport time scale arguments can be used to
  predict the abundance of CH$_4$ at its quenching point in the
  atmosphere.

Next, we examine the potential influence of the transport-induced quenching
on the hydrocarbon chemistry. Above the thermochemical regime in the
deep atmosphere, where equilibrium is maintained via rapid reaction
kinetics, a quenched regime may exist for some species, where rapid
atmospheric transport and slow reaction kinetics drive constituents
out of equilibrium.  As a result, the abundance of molecules can be
different from their equilibrium value at the same height in the
atmosphere.  We investigate  three pre-selected hydrocarbon species,
C$_2$H$_2$ , C$_2$H$_6$ and C$_2$H$_3$, which among the atmospheric
hydrocarbon inventory, become increasingly important in the deeper
atmospheric layers.

\sed{ There is good reason to think that CO and CH$_4$ are quenched, and
their quenching height may affect the dominant destruction pathway 
for C$_2$H$_x$ species. This
may happen in two ways. Either (1) CO or CH$_4$ directly destroys C$_2$H$_x$, or
(2) CO or CH$_4$ is involved in the formation or destruction of the species that destroy
C$_2$H$_x$. We are capable of applying our first-order approximation of
 $t_{\rm chem}$ to explore (1).  
Considering only the direct destruction of C$_2$H$_x$ is
a simplification that involves profound uncertainties. The full effect
of CH$_4$ and CO cannot be accounted for under such a simplification,
and these may impact the abundances of various species that destroy
C$_2$H$_x$ such that other destruction routes may dominate. This
would be the case, however, even if CH$_4$ and CO were not quenched; 
we only consider direct destruction routes for these C$_2$H$_x$ species.

The reactions that destroy C$_2$H$_2$, C$_2$H$_3$ and C$_2$H$_6$
 that involve CO or CH$_4$ and meet criteria (i) - (iii) in 
 Section \ref{ss:nonequapproach} are:
\begin{align}
{\rm C_2H_2 + CO} &\rightarrow {\rm C_2H + HCO}, \label{eqn:c2h2-quench}\\
{\rm C_2H_3 + CO} &\rightarrow {\rm C_3H_3O}, \label{eqn:c2h3-quench-1}\\
{\rm C_2H_3 + CH_4} &\rightarrow {\rm C_2H_4 + CH_3}, \label{eqn:c2h3-quench-2}\\
{\rm C_2H_6 + CO} &\rightarrow {\rm C_2H_5 + HCO}. \label{eqn:c2h6-quench}
\end{align}
with rate coefficients taken from \cite{Tsang1986}. There are no known
reactions involving C$_2$H$_6$ and CO/CH$_4$. In order to find the
maximum direct effect of CO and CH$_4$ quenching on relaxation
time-scales for C$_2$H$_2$, C$_2$H$_3$ and C$_2$H$_6$, we consider the
rates of these reactions with CO and CH$_4$ quenched at the highest
pressures considered in our} \textsc{Drift-Phoenix} \sed{ model
  atmospheres. It turns out that Reactions
  (\ref{eqn:c2h3-quench-1}),(\ref{eqn:c2h3-quench-2}) are both much
  slower than the destruction of C$_2$H$_3$ by H$_2$, and so the
  quenching of CO and CH$_4$ does not directly affect the quenching
  height for C$_2$H$_3$. The relaxation time-scale $t_{\rm chem}$ is
  shorter for both destruction reactions for C$_2$H$_2$ in Table
  \ref{table:reaction} than for Reaction \ref{eqn:c2h2-quench}, even
  assuming the maximum possible abundance of CO due to quenching.
  Reaction (\ref{eqn:c2h6-quench}) is endothermic, with a barrier of
  $\sim 43000$ K, but the reverse reaction is severely impeded in the
  upper atmosphere by the depletion of HCO. If CO is not quenched,
  then the reverse reaction, HCO + C$_2$H$_5$ dominates throughout the
  atmosphere.  If, however, CO is quenched at the highest pressure we
  consider for our model atmospheres, then Reaction
  (\ref{eqn:c2h6-quench}) dominates in the $\log g = 3$, solar
  metallicity case when $p_{\rm gas} \lesssim 10^{-5}$
  bar. Nevertheless, the relaxation time-scale for this reaction is
  much larger than any of the dynamical time-scales, and so this
  reaction does not help to determine the quenching height of
  C$_2$H$_6$.  The time-scale comparison for Reactions
  (\ref{eqn:c2h2-quench})-(\ref{eqn:c2h3-quench-2}) are plotted in
  Fig. \ref{fig:quench}.}

\begin{figure}
\includegraphics[scale=1.5]{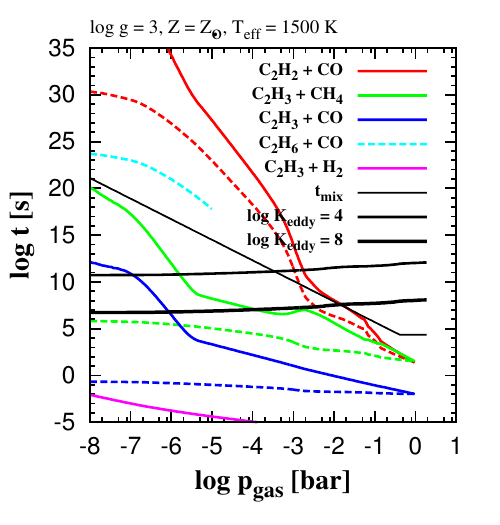}
\caption{Relaxation and dynamical time-scales for our $\log g = 3$, 
$T_{\rm eff} = 1500$ K, solar metallicity model atmosphere, as a function of
pressure [bar]. The colored lines represent relaxation time-scales for Reactions
(\ref{eqn:c2h2-quench}, red), (\ref{eqn:c2h3-quench-1},blue),
(\ref{eqn:c2h3-quench-2}, green) and (\ref{eqn:c2h6-quench}, cyan). Solid lines are the time-scales when CO and 
CH$_4$ are not quenched. Dashed lines are the time-scales when 
CO and CH$_4$ are quenched at $\sim 1$ bar. The plot for Reaction (\ref{eqn:c2h6-quench}) only
extends to $p_{\rm gas} < 10^{-5}$ bar, because at higher pressures, the reverse reaction, C$_5$H$_2$
+ HCO, dominates. The solid purple line is the relaxation time-scale for C$_2$H$_3 + $ H$_2$.}
\label{fig:quench}
\end{figure}

The reaction pathways studied here are approximated to be
rate-determining steps, since no data could be found available with
full reaction schemes of hydrocarbons in substellar environments. Each
molecular interconversion, such as CO/CH$_4$ and N$_2$/NH$_3$, are
usually a full reaction scheme that consists of many reactions
including the rate-determining step.  The applicability of reaction
schemes are also dependent on the atmospheric structure studied, and
the uncertainties on the kinetic rate coefficients make the field of
reaction kinetics very challenging as is clear from the wide
range of time-scales for single reactions in
Figs. \ref{fig:C2H2},\ref{fig:C2H6}.

For three pre-selected hydrocarbon species,
C$_2$H$_2$, C$_2$H$_6$ and C$_2$H$_3$, we find the following:

\sed{  C$_2$H$_2$:} Figure~\ref{fig:C2H2} compares
the chemical relaxation time-scales for C$_2$H$_2$ to the convective
mixing time-scale and turbulent diffusion time-scale. In the
solar-composition giant gas planet model, the quench level of
C$_2$H$_2$ \sed{ can occur at atmospheric pressures as high as $p_{\rm gas}
\approx 3 \times 10^{-3}$ bar down to $p_{\rm gas} \approx 10^{-5}$
bar. The intersection of $t_{\rm mix}$ and $t_{\rm chem}$
occurs at different pressures for all three models. When $\log g = 5$
(solar metallicity), quenching occurs somewhere in the range $10^{-3}$
bar $< p_{\rm gas} < 3 \times 10^{-2}$ bar. In the low
metallicity case ($\log g = 3$, [M/H]$ = -3$), quenching occurs within
the range $10^{-4}$ bar $< p_{\rm gas} < 3 \times 10^{-2}$ bar.}

\begin{figure*}
\includegraphics[scale=0.55,angle=-90]{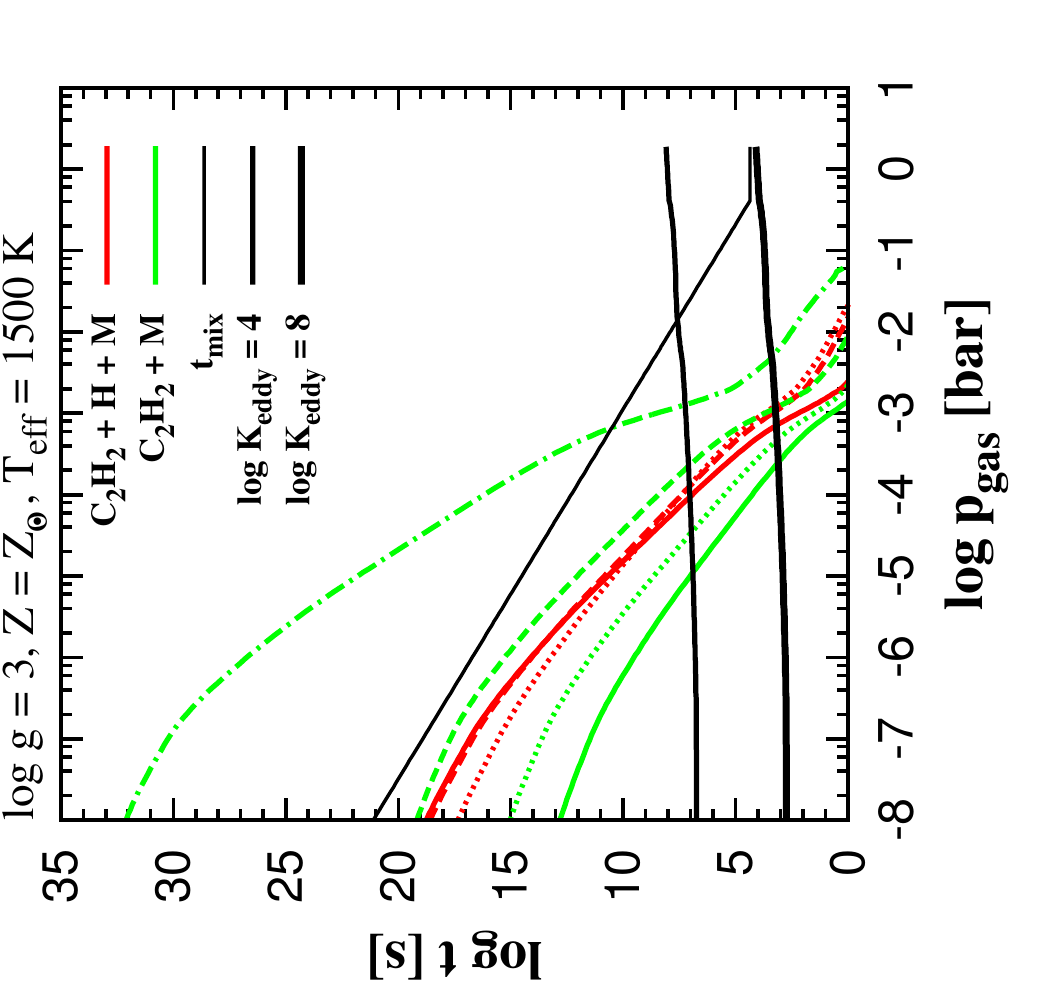}
\includegraphics[scale=0.55,angle=-90]{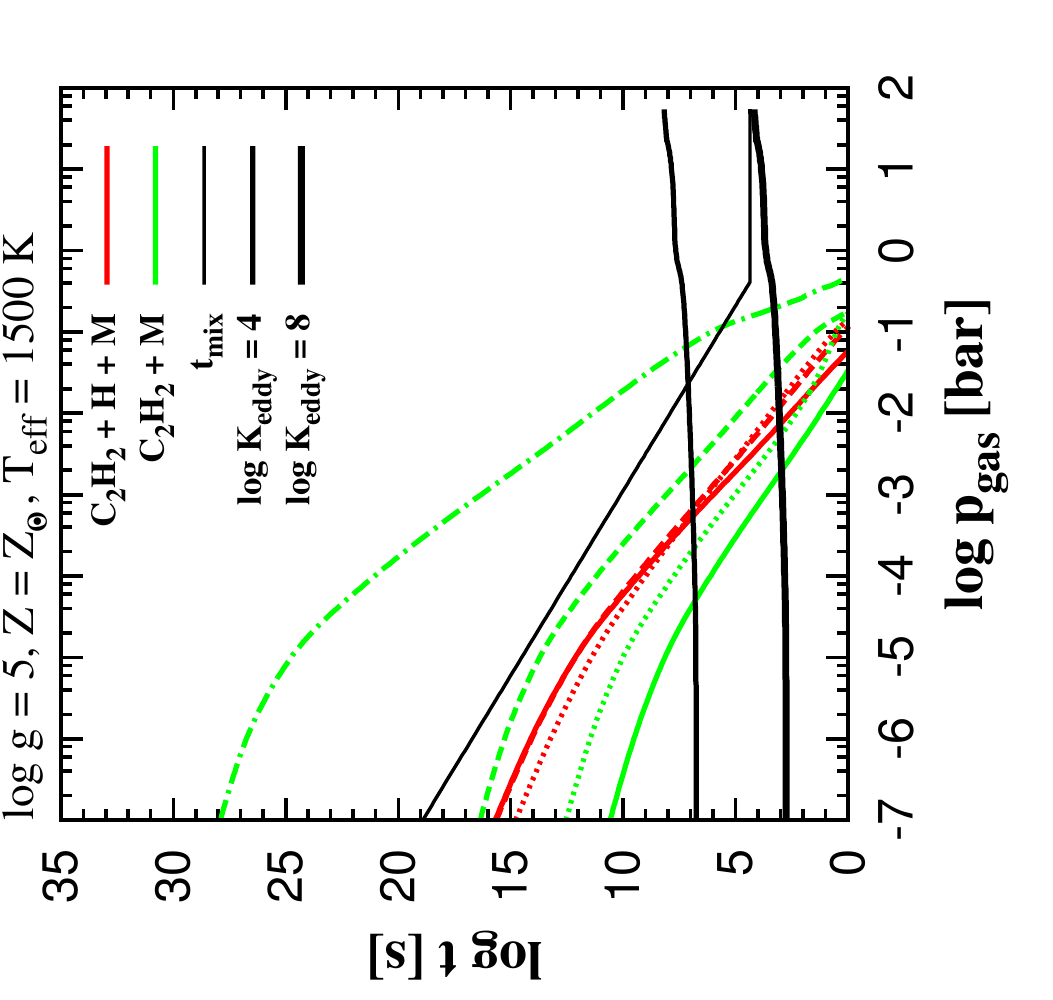}
\includegraphics[scale=0.55,angle=-90]{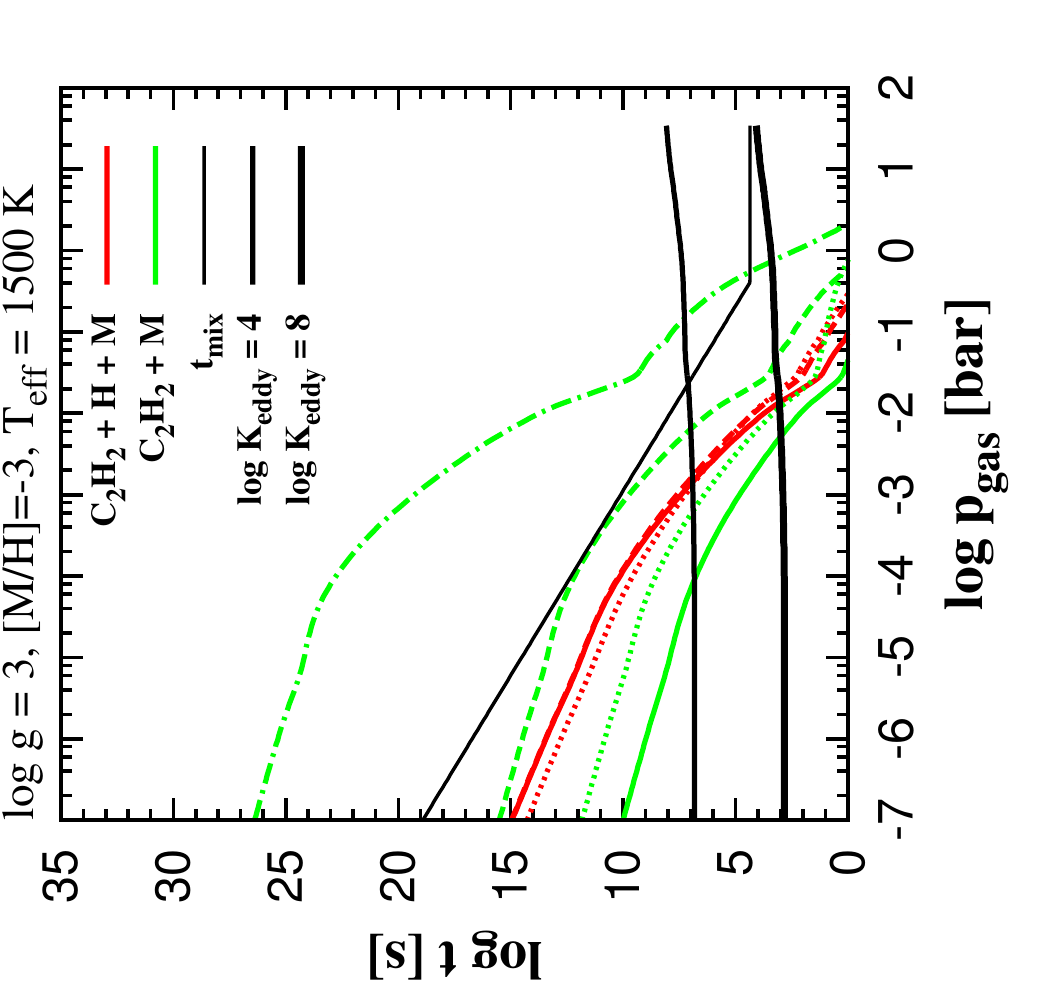}\\*[-0.1cm]
  \caption{Convective mixing timescale (\textit{thin black line}), the eddy diffusion mixing 
timescale (\textit{thick black lines}) and the reaction time for the destruction 
C$_2$H$_2$ when a) (left) $\log g = 3$, $Z = Z_{\sun}$, $T_{\rm eff} = 1500$ K (giant gas planet), 
b) (middle) $\log g = 5$, $Z = Z_{\sun}$, $T_{\rm eff} = 1500$ K (brown dwarf),
c) (right) $\log g = 3$, [M/H] = 0.3, $T_{\rm eff} = 1500$ K (oxygen-depleted giant gas planet). The
colored lines represent time-scales for various published rate coefficients. For the three-body destruction
pathway, we show rate coefficients from \citet[][solid red, also \citealt{Tsang1986,Hoyermann1968}]{Baulch1992}, 
\citet[][dashed red]{Tsang1986} and \citet[][dotted red]{Benson1967}. For the combustion reaction, coefficients
are from \citet[][solid green, also \citealt{Benson1989}]{Duran1989}, \citet[][dashed green]{Thraen1982}, \citet[][dotted green]{Palmer1964}
and \citet[][dash-dotted green, also \citealt{Tsang1986}]{Warnatz1984}.}
\label{fig:C2H2}
\end{figure*}

\begin{figure*}
\includegraphics[scale=0.55,angle=-90]{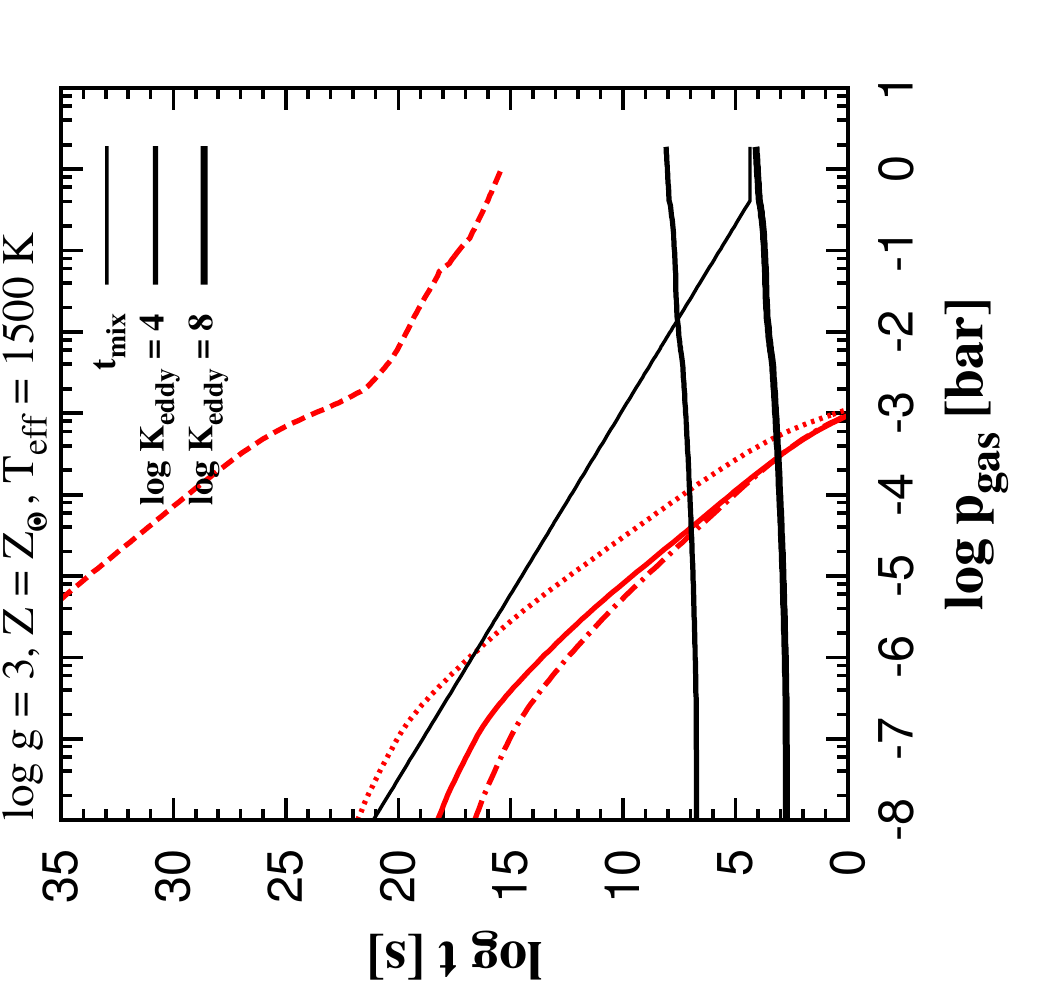}
\includegraphics[scale=0.55,angle=-90]{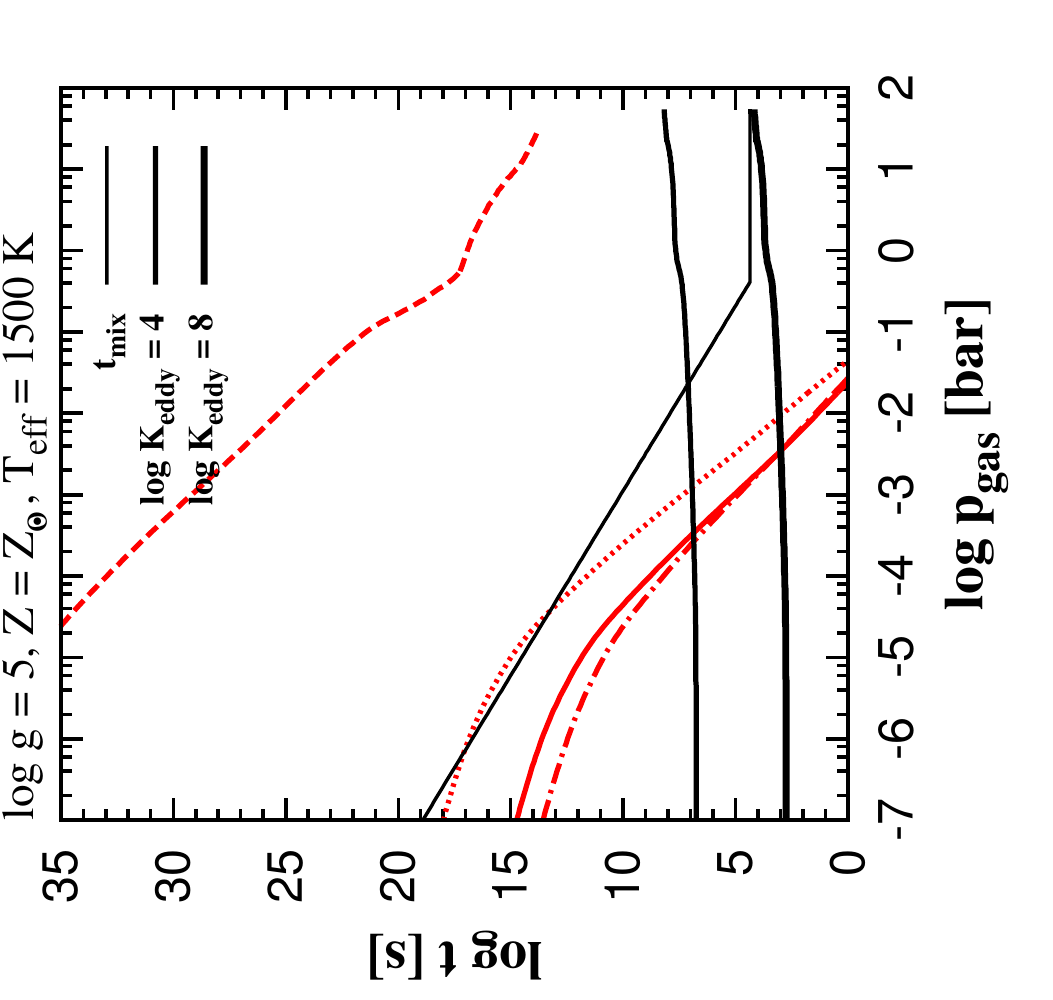}
\includegraphics[scale=0.55,angle=-90]{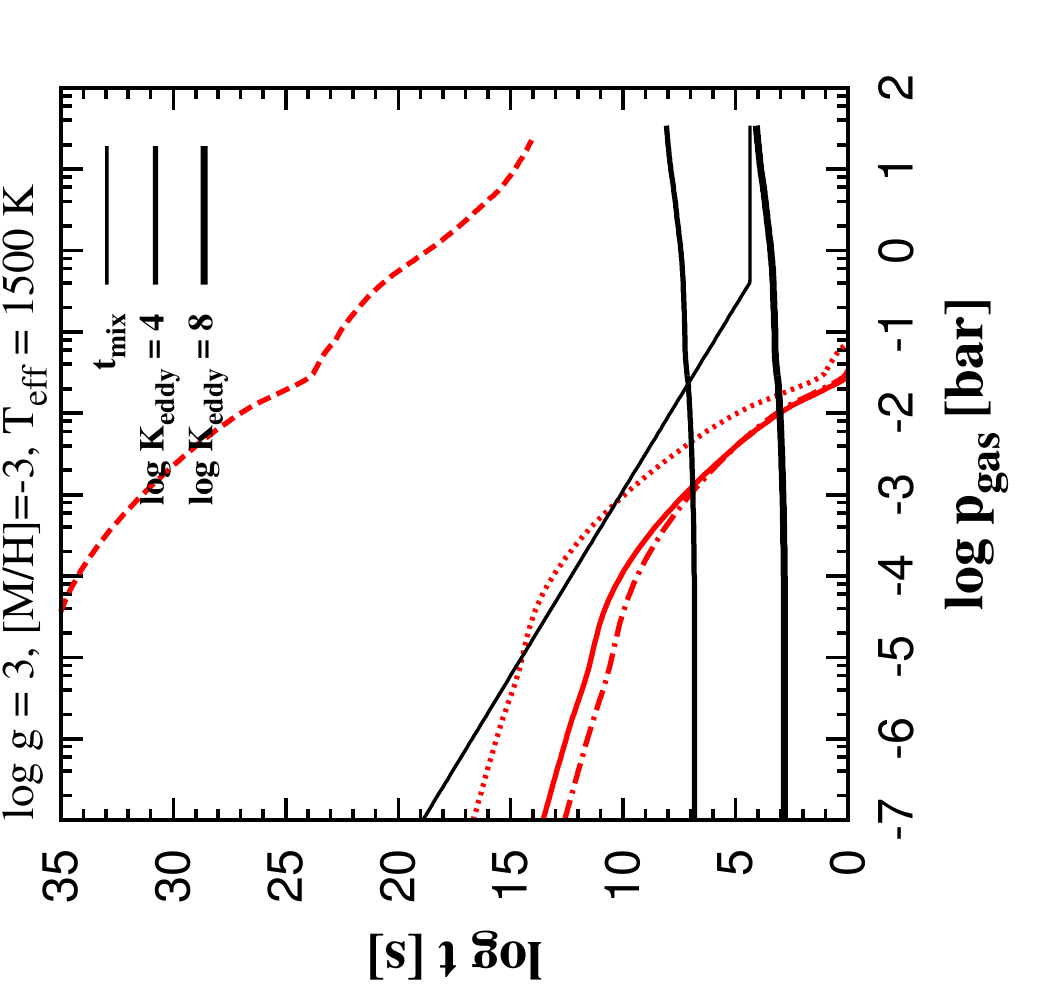}\\
\caption{Same as Figure \ref{fig:C2H2} but for C$_2$H$_6$ combustion. The time-scales for various rate-coefficients are shown in red. Rate coefficients
are taken from\citet[][solid]{Oehlschlaeger2005}, \citet[][dashed]{Baulch1992}, \citet[][dotted]{Izod1971} and \citet[][dash-dotted]{Warnatz1984}.
The time-scale resulting from the theoretical rate coefficient of \citet{Kiefer2005} were also considered, but the relaxation time-scale for their rate coefficient 
is greater than the age of the universe over the entire pressure range.}
\label{fig:C2H6}
\end{figure*}

Transport-induced quenching has been extensively studied also by \citet{Moses2011}
in the atmospheres of hot Jupiters. A dominant
interconversion scheme for C$_2$H$_2$ $\rightarrow$ CH$_4$ is proposed
\citep[][their Eq. 9]{Moses2011}, of which the rate-determining step
\begin{equation}
{\rm C_2H_2 + H + M} \longrightarrow {\rm C_2H_3 + M}
\end{equation}
is included in Table \ref{table:reaction}, along with the combustion
reaction for C$_2$H$_2$.
\citet{Moses2011} found that disequilibrium chemistry enhances the
abundances of acetylene (C$_2$H$_2$) in the atmospheres of hot Jupiter
exoplanets.

\sed{ C$_2$H$_3$:} The quenching-study approach fails for species that
have no quench point, i.e. no disequilibrium number density of the
species can be estimated for above the quench level, only a comparison
between the mixing and chemical timescales can be made.  
For example, the destruction of C$_2$H$_3$ through reaction with 
molecular hydrogen is unquenchable. 
\sed{ This can be
easily determined from Fig. \ref{fig:quench}, since the relaxation time-scale for
C$_2$H$_3$ is orders of magnitude lower than the fastest dynamical time-scales
considered for our model atmospheres.}
It results that vertical mixing cannot freeze out
C$_2$H$_3$ destruction, and one cannot use our simple approximation of
the quenching kinetics in this case to draw conclusions on the
disequilibrium abundance of C$_2$H$_3$.

\sed{ C$_2$H$_6$:}
\sed{ Figure \ref{fig:C2H6} shows our results for the
quenching of C$_2$H$_6$ according to our alternative method. 
The quench level candidates are intersection points
between the eddy diffusion timescales and $t_{\rm chem}$ for the
reaction:
\begin{equation}
{\rm C_2H_6 + M} \longrightarrow {\rm CH_3 + CH_3 + M}.
\label{eqn:c2h6}
\end{equation}
}For C$_2$H$_6$, quenching may
occur at pressures \sed{ as low as $10^{-3}$ bar}. By increasing the eddy
diffusion, the eddy diffusivity mixing time decreases. Consequently,
the intersection of $t_{\rm mix,eddy}$ and $t_{\rm chem}$ occurs at
a higher range of pressures yielding a higher non-equilibrium
number density of acetylene and ${\rm C_2H_6}$ over a large extent of
the atmosphere (compare thick black line in Figs.~\ref{fig:C2H2},
\ref{fig:C2H6}). \sed{ Despite the uncertainties in the rate 
coefficient for Reaction~\ref{eqn:c2h6}, a lower limit to the
quenching height can 
be established.}

\section{Discussion}
\label{sec:disc}

 Our exploration of the effects that dust formation has on gas-phase
metallicity and as a result on the equilibrium chemistry has yielded
some surprising results. Although only affecting the C/O by bringing it from
$0.5$ to $0.7$, dust formation results in the depletion of metals like silicon
and titanium by several orders of magnitude, and has had an orders-of-magnitude
impact on the chemistry. Also remarkable is the stability of benzine in the deep
atmosphere. We note particularly the high abundance of C$_4$H at low temperatures
and the impact of various published rate coefficients on predictions for the
quenching heights and time-scales, both of which we discuss below.

\subsection{C$_4$H abundance}
Our results suggest high abundances of C$_4$H which is somewhat
  surprising as C$_4$H is a radical and should therefor be destroyed
  easily. We checked the equilibrium constants and none of them
  behaves strangely when plotted, nor are they used outside that
  tested temperature interval. We cannot find any obvious errors
  with our calculation with respect to any of the
  molecules considered here, including C$_4$H.

Surprisingly high abundances of C$_4$H have been observed in the
  interstellar medium \citep[e.g.][]{Pety2005}, in cometary ice
  \citep{Geiss1999}, and are believed to play an important role in the
  complex carbon chemistry within Titan's atmosphere
  \citep{Berteloite2008}. It is believed that the high amount of
  C$_4$H cannot be accounted for by gas-phase thermochemistry, but is
  formed as a product of PAH destruction, within the cometary ice
  itself, or via photodissocation \citep{Leonori2008}. It is
  interesting that our thermochemical equilibrium contains high
  amounts of C$_4$H, without having accounted for any of these
  possible sources. The condensation of heavy elements in the gas onto
  grains, and the impact on the metallicity, may partially explain the
  high abundances of C$_4$H observed in comets and the interstellar
  medium, and inferred to be present in the atmosphere of Titan.


\subsection{Uncertainties due to differences in rate coefficient} 
\label{ss:uncertain}
Given the
various rate networks employed in the literature, we assess the impact
of the different material data on the quenching point. For this, we utilise
C$_2$H$_6$ quenching. The uncertainties in the rate coefficient for
Reaction (\ref{eqn:c2h6}) (Table~\ref{table:reaction}) span \sed{ an order of magnitude
or more amongst the different rate coefficient data sources. For example,
two values for $k_0$ [cm$^3$ s$^{-1}$] for the C$_2$H$_6$ combustion reaction
given in \citet{Baulch1992} have identical values of $n$ and $E_a$, but values of 
$A = 7.5 \times 10^{-20}$ cm$^3$ s$^{-1}$ and $4.5 \times 10^4$ cm$^3$ s$^{-1}$,
a difference of about 24 orders of magnitude. A more typical disagreement would
be between the $k_0$ values for this same reaction between 
\citet{Baulch1992} and \citet{Warnatz1984}, which at 800 K is about an order of
magnitude, although these two rate coefficients come into much better agreement at
high temperatures.}

 Concerning the destruction of C$_2$H$_2$ by the three-body
 interaction with hydrogen, we examined in detail published rate
 coefficients from reviews \citep{Baulch1992,Tsang1986}, experiment
 \citep{Hoyermann1968} and theory \citep{Benson1967}. For the
 combustion of C$_2$H$_2$, we examined the review of rate coefficients
 from \citet{Duran1989}, as well as the experimentally determined rate
 coefficients from \citet{Thraen1982} and \citet{Palmer1964} and the
 rate coefficients determined theoretically by
 \citet{Benson1989}. Finally, for C$_2$H$_6$ combustion into 2CH$_3$,
 we considered the reviewed rates from \citet{Baulch1992,Warnatz1984},
 as well as experimental \citep{Oehlschlaeger2005,Izod1971} and
 theoretical calculations \citep{Kiefer2005} of these rate
 coefficients. The range of values for the rate coefficients over the
 pressure range is incredible, and its impact on the chemical
 time-scales spans 15 orders of magnitude, as can be seen in
 Figs. \ref{fig:C2H2} and \ref{fig:C2H6}.  Three-body rate
 coefficients are very difficult to constrain from high temperature
 experiments, and theoretical work is therefore also fairly
 unconstrained. These large uncertainties pose a problem not only for
 our work, but  even more so for the non-linear chemical kinetics models
 applied to these atmospheres. Improved experimental and theoretical
 determinations of these rate constants are essential  to
 progress beyond the first order approximation employed in this paper.

This leads in the case of C$_2$H$_6$ that there is no
definite upper limit to the quenching pressure.  Figure~\ref{fig:C2H6}
depicts the results for the different rate coefficients in different
lines styles which shows that it is possible that $t_{\rm chem} >
t_{\rm diff}$ for the entire range of pressures considered in the
\textsc{Drift-Phoenix} model atmospheres.  The reason for this is the
uncertainty of the values of the rate coefficients alone.  It will be
important for future chemical kinetics modelling to more carefully
explore the efficiency of Reaction (\ref{eqn:c2h6}).

\sed{ In the case of C$_2$H$_2$, the termolecular destruction with
  atomic hydrogen has a relatively small uncertainty, and if it were
  the only dominant reaction for destroying C$_2$H$_2$, its quenching
  height would likewise be well-constrained. The rate coefficient for
  C$_2$H$_2$ combustion is far less accurate, but it still provides a
  range of quenching heights; the uncertainties do not overwhelm our
  analysis. The fact that C$_2$H$_3$ reacts with molecular hydrogen
  means that it should not have a quenching height at all. 

We note again that the data uncertainties do also apply to
full-network considerations. Every quenching height given in the
literature will therefore change if the material data change. Hence,
quenching heights should in general be rather given as a limit~$\pm$~uncertainty.}

\section{Conclusion}
It must be acknowledged that small hydrocarbon molecules are able to
form in an oxygen-rich environment such as the atmosphere of Brown
Dwarfs and giant gas planets. These molecules do not form in very
significant concentrations in comparison to carbon-rich atmospheres;
nonetheless, an increased surface gravity and/or decreasing
metallicity combined with a greater C/O ratio inside the dust clouds
improve the chance of PAH formation.  A decrease of the oxygen
abundance caused by oxygen-depletion due to cloud formation does
support the appearance of complex carbon-binding molecules. These
results contradict the general belief that hydrocarbon equilibrium
chemistry is not expected in the atmospheres of Brown Dwarfs and giant
gas planets where the C/O ratio is less than unity.  It must be noted
that hydrocarbon chemistry in irradiated giant planets, through
non-equilibrium photochemistry, is theoretically predicted and
observed (Zahnle et al. 2009).  The formation of hydrocarbons in hot
Jupiters with temperatures below 1000K is driven by the
photodissociation of methane; the products - C$_2$H$_2$, C$_2$H$_4$
and C$_2$H$_6$ - further polymerise to build complex PAHs and
hydrocarbon aerosols, called soots, which are thought to be involved
in the prebiotic evolutionary processes towards the emergence of amino
acids \citep{Tielens2008}.  For wavelengths at which the dust cloud is
transparent, the deep atmospheric layers can be observed. Due to the
relatively significant number densities of benzene (C$_6$H$_6$), the
vinyl radical (C$_2$H$_3$) and acetylene (C$_2$H$_2$) predicted in our
work in this region, one might expect a signature in the absorption
lines.

A recent work by \citet{Fortney2012} discussed the possibility of the
formation of carbon-rich giant planets in disks where the
``condensation of solids can lead to non-stellar C/O ratios in nebula
gases'', in accordance with the idea used in the present
work. Furthermore, \citet{Fortney2012} raises the question of the detection
of carbon-rich Brown Dwarfs that may have been eluded so far ($2$MASS
and SDSS) simply because the spectral appearance of a Brown Dwarf with
refractory clouds that remove oxygen from the gas-phase will be
different: different molecules will influence the opacities and thus,
the absorption lines in the atmosphere. The present work is a step forward in
determining the chemical species whose opacities may yield
to spectra that appear distinctly different from objects with no
oxygen-depleted refractory clouds.

An interesting process to consider would be the transport of the
deeper-layer hydrocarbons upward into the cloud layer. Additionally,
one could imagine hydrocarbons sticking on the surface of dust grains,
producing dark soot grains. This new piece of chemistry could lower
the albedo of a gas giant or a Brown Dwarf, by rendering its spectral
appearance ``darker''. A recent work by \citet{Tian2012} studied the
formation mechanism of PAH molecules in interstellar and circumstellar
environments by looking at reactions of acetylene over silicate
particles like forsterite (MgSiO$_4$), a particularly abundant dust
particle in the clouds of brown dwarfs. Their experiments lead to the
production of gas-phase PAHs such anthracene, naphthalene,
phenanthrene and pyrene.

\section{Acknowledgements}
We highlight financial support of the European Community under the FP7
by an ERC starting grant.  The computer support at the School of
Physics \& Astronomy in St Andrews is highly acknowledged.  Most
literature serach has been performed using ADS. Our local computer
support is highly acknowledged.

\label{lastpage}

\end{document}